\documentclass[pre]{revtex4}

\usepackage{graphicx}
\usepackage{dcolumn}
\usepackage{bm}
\usepackage{subfigure}
\usepackage{amssymb}
\usepackage{graphics}
\usepackage{epsfig}
\newcommand{\thickhline}{\noalign{\hrule height 0.8pt}}
\usepackage{threeparttable}

\begin{document}


\title{A new effective field theory for spin-$S$ $(S\leq1)$ dilute Ising ferromagnets}
\author{\"{U}mit Ak{\i}nc{\i}}
\author{Yusuf Y\"{u}ksel}
\author{Hamza Polat}\email{hamza.polat@deu.edu.tr; Phone: +90 2324128672; fax: +90 2324534188.}
\affiliation{ Department of Physics, Dokuz Eyl\"{u}l University,
TR-35160 Izmir, Turkey}
\date{\today}

\begin{abstract}
Site diluted spin-1/2 Ising and spin-1 Blume Capel (BC) models in the presence of transverse field interactions are examined by introducing an effective-field approximation that takes into account the multi-site correlations in the cluster of a considered lattice with an improved configurational averaging technique. The critical concentration below which the transition temperature reduces to zero is determined for both models, and the estimated values are compared with those obtained by the other methods in the literature. It is found that diluting the lattice sites by non magnetic atoms may cause some drastic changes on some of the characteristic features of the model. Particular attention has been paid on the global phase diagrams of a spin-1 BC model, and it has also been shown that the conditions for the occurrence of a second order reentrance in the system is rather complicated, since the existence or extinction of reentrance is rather sensitive to the competing effects between $D/J$, $\Omega/J$ and $c$.
\end{abstract}



\maketitle

\tableofcontents

\section{Introduction}\label{intro}
Investigation of disorder effects on the critical phenomena has a long history and there have been a great many of theoretical studies focused on disordered magnetic materials with quenched randomness where the random variables of a magnetic system such as random fields \cite{larkin,imry} or random bonds \cite{edwards,sherington} may not change its value over time. On the other hand, site diluted ferromagnets constitute another example of magnetic systems with quenched disorder such as a compound $\mathrm{A_{x}B_{1-x}C}$  where magnetic $\mathrm{A}$ atoms in a pure magnet $\mathrm{AC}$ are replaced by non-magnetic $\mathrm{B}$ impurities. Formerly, Sato et al. \cite{sato} have shown that in a dilute lattice a Curie or a N\'{e}el temperature does not appear until a finite concentration of magnetic atoms is obtained if the atomic distribution is random. They have also found that this concentration depends on the coordination number of the lattice. After this seminal work of Sato et al. \cite{sato}, much attention has been paid to site dilution problem and the situation has been handled by a wide variety of techniques such as Bethe-Peierls-Weiss (BPW) method \cite{charap}, renormalization group (RG) technique \cite{yeomans1,yeomans2,benyoussef,mina}, correlated effective field theory (CEFT) \cite{taggart,kaneyoshi1,bobak1}, effective field theory (EFT) based on decoupling (or Zernike \cite{zernike}) approximation (DA) \cite{Alcantara,boccara,kaneyoshi2,fittipaldi,balcerzak1,li,kaneyoshi3,yang,tucker,saber1,bobak2,tucker1,kaneyoshi4,saber2,sarmento,kaneyoshi5,kaneyoshi6,kaneyoshi7,liang}, an integral representation method \cite{mielnicki}, Monte Carlo (MC) simulation technique \cite{marro,kim,neda,ballesteros}, Bogoliubov inequality approach \cite{alcazar}, Bethe-Peierls approximation (BPA) \cite{labarta}, finite cluster approximation (FCA) which gives results identical to those obtained by EFT for a one spin cluster \cite{bobak3,wiatrowski,mockovciak,bobak4}, third order Matsudaira approximation \cite{balcerzak2}, EFT with probability distribution technique \cite{kerouad1,bakkali,tucker2,kerouad2,saber3,htoutou1,htoutou2} and cluster variational method (CVM) \cite{balcerzak3}. Among the theoretical works mentioned above, some of the authors extended the standard dilution problem to more complicated versions by taking into account the transverse field interactions \cite{degennes}, random fields and random bonds, as well as bilinear and biquadratic exchange couplings and crystal field interactions for the systems with $S>1/2$.

Mean field theory (MFT) of site dilution problem predicts that the system always has a finite critical temperature and stays in a ferromagnetic state at lower temperatures, except that $c=0$ where $c$ denotes the magnetic atom concentration. Therefore, it is not capable of locating a critical site concentration at which the transition temperature reduces to zero. The reason is due to the fact that MFT neglects single-site and multi-spin correlations. On the other hand, EFT based on DA accounts all the single site correlations, but it also neglects multi-spin correlations between different sites. Hence, EFT provides results that are superior to those obtained within the traditional MFT. Furthermore, CEFT which is an extension of EFT partially takes into account the effects of multi-spin correlations and improves the results of conventional EFT in many cases. Based on the physical aspects of the problem, whether in EFT or CEFT formalism, evaluation of configurational averages emerging in definition of spin identities plays a critical role. However, as mentioned by Tucker \cite{tucker}, the conventional configurational averaging technique applied in Refs. \cite{taggart,kaneyoshi1,boccara,yang,kaneyoshi4,kaneyoshi6,kaneyoshi7} is based on a procedure that decouples the site occupation variable from the thermal average of spin variables, even when both quantities referred to the same site while in Refs. \cite{balcerzak1,tucker,saber1,bobak2,tucker1,sarmento,bobak4}, the authors used an improved configurational averaging method in which only the correlations between quantities pertaining to different sites are neglected (decoupled). However, it is possible to improve the accuracy of these methods by including multi-site, as well as single site correlations.

In this paper, we describe a new type of EFT method for investigating the thermal and magnetic properties of a site diluted Ising model on 2D lattices. Recently, we have successfully applied our method to random bond \cite{akýncý1} and random field \cite{akýncý2} problems on 2D and 3D lattices with various coordination numbers. As we emphasized in these previous works, an advantage of the approximation method proposed by this method is that no decoupling procedure is used for the higher-order correlation functions. Therefore, it is expected that the accuracy of the results obtained within the present work may improve those of the works based on conventional and improved DA. For this purpose, we organized the paper as follows: In Sec. \ref{formulation} we briefly present the formulations. The results and discussions are presented in Sec. \ref{results}, and finally Sec. \ref{conclude} contains our conclusions.

\section{Formulation}\label{formulation}
In this section, we give the formulation of the present study for site diluted spin-1/2 Ising and spin-1 Blume Capel (BC) models on 2D lattices. As our model, we consider $N$ identical spins arranged on a 2D regular lattice. Then we define a cluster on the lattice which consists of a central spin labeled $S_{0}$ and $q$ perimeter spins being the nearest neighbors of the central spin. The cluster consists of $(q+1)$ spins being independent from the value of $S$. The nearest-neighbor spins are in an effective field produced by the outer spins, which can be determined by the condition that the thermal average of the central spin is equal to that of its nearest-neighbor spins. In the following subsections, we give a detailed discussion of how the present method can be formulated for spin-1/2 Ising and spin-1 BC models with quenched site dilution.
\subsection{Site diluted spin-$\frac{1}{2}$ system}\label{formualtion1}
As a site diluted spin-$1/2$ Ising model,  we consider the following Hamiltonian
\begin{equation}\label{eq1}
H=-J\sum_{<i,j>}c_{i}c_{j}S_{i}^{z}S_{j}^{z},
\end{equation}
where the summation is over the nearest-neighbor pairs of spins and the operator $S_{i}^{z}$ takes the values $S_{i}^{z}=\pm1$. We assume that the lattice sites are randomly diluted and $c_{i}$ denotes a site occupation variable which equals to $1$ if the site is occupied by a magnetic atom  or to $0$ if it is empty.

According to the Callen identity \cite{callen} for the spin-1/2 Ising  system, the thermal average of the identity $c_{i}S_{i}^{z}$ at the site $i$ is given by
\begin{equation}\label{eq2}
c_{i}\left\langle \{f_{i}\}S_{i}^{z}\right\rangle=c_{i}\left\langle \{f_{i}\} \tanh\left[\beta c_{i}\left(J\sum_{j}c_{j}S_{j}\right)\right] \right\rangle,
\end{equation}
where $\beta=1/k_{B}T$, $j$ expresses the nearest-neighbor sites of the central spin and $\{f_{i}\}$ can be any function of the Ising variables as long as it is not a function of the site. Applying the differential operator technique \cite{honmura,kaneyoshi8} in Eq. (\ref{eq2}) and using the relation
\begin{equation}\label{eq3}
\exp(\alpha c_{i})=c_{i}\exp(\alpha)+1-c_{i},
\end{equation}
with the fact that $c_{i}^{n}=c_{i}$, we get
\begin{equation}\label{eq4}
c_{i}\left\langle \{f_{i}\}S_{i}^{z}\right\rangle=c_{i}\left\langle \{f_{i}\} \prod_{j=1}^{q}\exp\left(Jc_{j}S_{j}^{z}\nabla\right) \right\rangle \tanh(\beta x)|_{x=0}.
\end{equation}
By putting Eq. (\ref{eq3}) into Eq. (\ref{eq4}) we obtain
\begin{equation}\label{eq5}
c_{i}\left\langle \{f_{i}\}S_{i}^{z}\right\rangle=c_{i}\left\langle \{f_{i}\}\prod_{j=1}^{q}\left\{c_{j}\cosh(J\nabla)+c_{j}S_{j}^{z}\sinh(J\nabla)+1-c_{j}\right\} \right\rangle \tanh(\beta x)|_{x=0},
\end{equation}
where $\nabla$ is a differential operator, $q$ is the coordination number of the lattice, and $\langle ...\rangle$ represents the thermal average. Eq. (\ref{eq5}) is valid only for a given specific magnetic atom configuration. Hence, if we consider configurational averages then we may rewrite Eq. (\ref{eq5}) as
\begin{equation}\label{eq6}
\left\langle c_{i}\left\langle \{f_{i}\}S_{i}^{z}\right\rangle\right\rangle_{r}=\left\langle c_{i}\left\langle \{f_{i}\} \prod_{j=1}^{q}\left\{c_{j}\cosh(J\nabla)+c_{j}S_{j}^{z}\sinh(J\nabla)+1-c_{j}\right\} \right\rangle \right\rangle_{r} \tanh(\beta x)|_{x=0},
\end{equation}
where $\langle...\rangle_{r}$ represents random configurational averages. When the right-hand side of Eq. (\ref{eq6}) is expanded, the multi-site correlation functions appear. The simplest approximation, and one of the most frequently adopted is to decouple these correlations which is called decoupling approximation (DA). In conventional manner, eliminating the term $c_{i}$ from both sides of Eq. (\ref{eq5}) then performing the configurational average with $\{f_{i}\}=1$ leads to the following equation,
\begin{equation}\label{eq7}
\left\langle\left\langle S_{i}^{z}\right\rangle\right\rangle_{r}=\left\langle\left\langle \prod_{j=1}^{q}\left\{c_{j}\cosh(J\nabla)+c_{j}S_{j}^{z}\sinh(J\nabla)+1-c_{j}\right\} \right\rangle\right\rangle_{r} \tanh(\beta x)|_{x=0}.
\end{equation}
In conventional DA one expands the right-hand side of Eq. (\ref{eq7}) then decouples the multi-site correlations according to
\begin{equation}\label{eq8}
\left\langle\left\langle c_{i}...c_{j}c_{k}S_{k}^{z}c_{l}S_{l}^{z}...c_{m}S_{m}^{z}\right\rangle\right\rangle_{r}\cong\left\langle c_{i}\right\rangle_{r}...\left\langle c_{j}\right\rangle_{r}\left\langle c_{k}\right\rangle_{r}\left\langle\left\langle S_{k}^{z}\right\rangle\right\rangle_{r}\left\langle c_{l}\right\rangle_{r}\left\langle\left\langle S_{l}^{z}\right\rangle\right\rangle_{r}...\left\langle c_{m}\right\rangle_{r}\left\langle\left\langle S_{m}^{z}\right\rangle\right\rangle_{r}
\end{equation}
with
\begin{eqnarray}\nonumber
\left\langle c_{\alpha}\right\rangle_{r}=c \quad \mathrm{and} \quad \left\langle\left\langle S_{\alpha}^{z}\right\rangle\right\rangle_{r}=m \qquad \alpha=i,...j,k,l,...,m.
\end{eqnarray}
However, this approximation decouples the site occupation variable from the thermal and configurational averages of spin variable, even when both quantities referred to the same site.

On the other hand, an improved version of decoupling approximation deals  with the quantity $\left\langle c_{i}\left\langle S_{i}^{z}\right\rangle\right\rangle_{r}$. In other words, in an improved decoupling procedure, one expands the right-hand side of Eq. (\ref{eq6}) instead of Eq. (\ref{eq7}) and decouples the multi-site correlations according to
\begin{equation}\label{eq9}
\left\langle\left\langle c_{i}...c_{j}c_{k}S_{k}^{z}c_{l}S_{l}^{z}...c_{m}S_{m}^{z}\right\rangle\right\rangle_{r}\cong\langle c_{i}\rangle_{r}...\langle c_{j}\rangle_{r}\left\langle c_{k}\left\langle S_{k}^{z}\right\rangle\right\rangle_{r}\left\langle c_{l}\left\langle S_{l}^{z}\right\rangle\right\rangle_{r}...\left\langle c_{m}\left\langle S_{m}^{z}\right\rangle\right\rangle_{r}
\end{equation}
with
\begin{eqnarray}\nonumber
\left\langle c_{i}\right\rangle_{r}=\left\langle c_{j}\right\rangle_{r}=c \quad \mathrm{and} \quad \left\langle c_{\alpha}\left\langle S_{\alpha}^{z}\right\rangle\right\rangle_{r}=m \qquad \alpha=k,l,...,m
\end{eqnarray}
In this approximation, only the correlations between quantities pertaining to different sites are neglected. A detailed discussion about these configurational averaging techniques is also given by Tucker \cite{tucker}. Whether conventional method or improved one, the papers which utilize these approximations claim that if we try to treat exactly all the spin-spin correlations emerging on the right-hand side of Eqs. (\ref{eq6}) and (\ref{eq7}), the problem becomes mathematically intractable. In order to overcome this point, recently we proposed an approximation that takes into account the correlations between different sites in the cluster of a considered lattice \cite{akýncý1,akýncý2}. Namely, an advantage of the approximation method proposed by those studies is that no decoupling procedure is used for the higher-order correlation functions.

We state that hereafter, we will carry on the formulation of the dilute spin-$1/2$ system for a honeycomb lattice $(q=3)$, however a brief explanation of the method for a square lattice $(q=4)$ can be found in \ref{appendixa}. Now, if we expand the right-hand side of Eq. (\ref{eq6}) for $q=3$ \textit{without using DA}, we get some certain identities in the form
\begin{equation}\label{eq10}
\left\langle\left\langle c_{i}...c_{j}c_{k}S_{k}^{z}c_{l}S_{l}^{z}...c_{m}S_{m}^{z}\right\rangle\right\rangle_{r}=\left\langle c_{i}...c_{j}\right\rangle_{r}\left\langle\left\langle c_{k}S_{k}^{z}c_{l}S_{l}^{z}...c_{m}S_{m}^{z}\right\rangle\right\rangle_{r}.
\end{equation}
In Eq. (\ref{eq10}), we use the fact that occupation number $c_{i}$ of a given site $i$ is independent from the thermal average, as long as the correlation function does not contain a spin variable $S_{i}^{z}$, and the site occupation numbers pertaining to different sites are assumed to be statistically independent from each other. Hence, we may rearrange Eq. (\ref{eq10}) as
\begin{equation}\label{eq11}
\left\langle\left\langle c_{i}...c_{j}c_{k}S_{k}^{z}c_{l}S_{l}^{z}...c_{m}S_{m}^{z}\right\rangle\right\rangle_{r}=\left\langle c_{i}\right\rangle_{r}...\left\langle c_{j}\right\rangle_{r}\left\langle\left\langle c_{k}S_{k}^{z}c_{l}S_{l}^{z}...c_{m}S_{m}^{z}\right\rangle\right\rangle_{r}.
\end{equation}
where $\left\langle c_{i}\right\rangle_{r}=\left\langle c_{j}\right\rangle_{r}=c$. In the present formulation, it is clear that Eq. (\ref{eq11}) improves EFT based on Eqs. (\ref{eq8}) and (\ref{eq9}) by taking into account the multi-site correlations. With the help of Eq. (\ref{eq11}), and by expanding the right-hand side of Eq. (\ref{eq6}) for the central site $c_{0}S_{0}^{z}$ with $\{f_{i}\}=1$ we have
\begin{eqnarray}\label{eq12}
m=\left\langle\left\langle c_{0}S_{0}\right\rangle\right\rangle_{r}=x_{1}=(3c-6c^{2}+3c^{3})x_{4}K_{1}+(6c^{2}-6c^{3})x_{4}K_{2}+3c^{3}x_{4}K_{3}+cx_{6}K_{4}.
\end{eqnarray}
where the terms $x_{i}$ in Eq. (\ref{eq12}) are defined in \ref{appendixa}. In obtaining Eq. (\ref{eq12}) we use the fact that $\tanh(\beta x)$ is an odd function. Hence, only the odd coefficients give non-zero contribution which can be given as follows:
\begin{eqnarray}\label{eq13}
\nonumber
K_{1}&=&\sinh(J\nabla)\tanh(\beta x)|_{x=0},\\
\nonumber
K_{2}&=&\cosh(J\nabla)\sinh(J\nabla)\tanh(\beta x)|_{x=0},\\
\nonumber
K_{3}&=&\cosh^{2}(J\nabla)\sinh(J\nabla)\tanh(\beta x)|_{x=0},\\
K_{4}&=&\sinh^{3}(J\nabla)\tanh(\beta x)|_{x=0}.
\end{eqnarray}
For comparison, if we apply the improved decoupling approximation given in Eq. (\ref{eq9}) then Eq. (\ref{eq12}) reduces to
\begin{equation}\label{eq14}
m=(3c-6c^{2}+3c^{3})mK_{1}+(6c^{2}-6c^{3})mK_{2}+3c^{3}mK_{3}+cm^{3}K_{4},
\end{equation}
which is identical to those obtained in Refs. \cite{balcerzak1,tucker,bobak2}. Additionally, applying the conventional method (\ref{eq8}) gives the following result
\begin{equation}\label{eq14b}
m=(3c-6c^{2}+3c^{3})mK_{1}+(6c^{2}-6c^{3})mK_{2}+3c^{3}mK_{3}+c^{3}m^{3}K_{4}.
\end{equation}
It seems like it is fortuitous that although, the equations of states of approximations (\ref{eq8}) and (\ref{eq9}) are differ from each other in the last term, they give the same phase diagram in $(k_{B}T_{c}/J-c)$ plane. The reason comes from the fact that both approximations ignore the term $m^{3}$ in the limit $T\rightarrow T_{c}$. Hence, it should be emphasized that the importance and distinction of our method becomes evident by expansion of Eq. (\ref{eq6}) without using any kind of DA.

The next step is to carry out the configurational and thermal averages of the perimeter site in the system, and it is found as
\begin{eqnarray}\label{eq15}
\left\langle\left\langle \{f_{\delta}\}c_{\delta}S_{\delta}\right\rangle\right\rangle_{r}=\left\langle c_{\delta} \left\langle \{f_{\delta}\}\left(c_{0}\cosh(J\nabla)+c_{0}S_{0}\sinh(J\nabla)+1-c_{0}\right)\right\rangle\right\rangle_{r}\tanh(\beta(x+\gamma)).
\end{eqnarray}
From Eq. (\ref{eq15}) with $\delta=\{f_{\delta}\}=1$ we get the following identity
\begin{eqnarray}\label{eq16}
\left\langle\left\langle c_{1}S_{1}\right\rangle\right\rangle_{r}=x_{4}=(c-c^{2})A_{1}+c^{2}A_{2}+cx_{1}A_{3}.
\end{eqnarray}
For the sake of simplicity, the superscript $z$ is omitted from the left- and right-hand sides of Eqs. (\ref{eq12}) and (\ref{eq16}). The coefficients
in Eq. (\ref{eq16}) are given as
\begin{eqnarray}\label{eq17}
\nonumber
A_{1}&=&\tanh(\beta(x+\gamma))|_{x=0},\\
\nonumber
A_{2}&=&\cosh(J\nabla)\tanh(\beta(x+\gamma))|_{x=0},\\
A_{3}&=&\sinh(J\nabla)\tanh(\beta(x+\gamma))|_{x=0}.
\end{eqnarray}
The coefficients in Eqs. (\ref{eq13}) and (\ref{eq17}) can easily be calculated by applying a mathematical relation, $e^{\alpha\nabla}f(x)=f(x+\alpha)$.  In Eq. (\ref{eq17}), $\gamma=(q-1)A$ is the effective field produced by the $(q-1)$ spins outside of the cluster, and $A$ is an unknown parameter to be determined self-consistently.

Eqs. (\ref{eq12}) and (\ref{eq16}) are the fundamental correlation functions of the system. On the other hand, for a honeycomb lattice, taking Eqs. (\ref{eq6}) and (\ref{eq15}) as basis, we derive a set of linear equations of the site correlation functions in the system. At this point, we assume that (i) the correlations depend only on the distance between the spins and (ii) the average values of a central site and its nearest-neighbor site (it is labeled as the perimeter site) are equal to each other with the fact that, in the matrix representations of spin operator $\hat{S}$, the spin-1/2 system has the property $(\hat{S})^{2}=1$. Thus, the number of linear equations obtained for $q=3$ and $q=4$ reduces to six and eight, respectively, and the complete sets are given in \ref{appendixa}.

Finally, e.g. if Eq. (\ref{eq2a}) for $q=3$ is written in the form of $6\times6$ matrix and solved in terms of the variables $x_{i}$ $(i=1,2,...6)$ of the linear equations, all of the site correlation functions can be easily determined as functions of the temperature and Hamiltonian parameters. Since the thermal and configurational average of the central site is equal to that of its nearest-neighbor sites within the present method, the unknown parameter $A$ can be numerically determined by the relation
\begin{equation}\label{eq18}
x_{1}=x_{4}.
\end{equation}
By solving Eq. (\ref{eq18}) numerically for a given fixed set of Hamiltonian parameters, we obtain the parameter $A$. Then we use the numerical values of $A$ to obtain the site correlation functions which can be found from Eq. (\ref{eq2a}). Note that $A=0$ is always a root of Eq. (\ref{eq18}) corresponding to the disordered state of the system whereas the nonzero root of $A$ in Eq. (\ref{eq18}) corresponds to the long-range-ordered state of the system. Once the site correlation functions have been evaluated then we can give the numerical results for the thermal and magnetic properties of the system. Since the effective field $\gamma$ is very small in the vicinity of $k_{B}T_{c}/J$, we can obtain the critical temperature for the fixed set of Hamiltonian parameters by solving Eq. (\ref{eq18}) in the limit of $\gamma\rightarrow0$, and then we can construct the whole phase diagrams of the system.

\subsection{Site diluted spin-$1$ Blume-Capel model with transverse field interactions}\label{formualtion2}
Site diluted spin-1 Blume-Capel (BC) \cite{blume,capel} model with transverse field interaction is represented by the following Hamiltonian
\begin{equation}\label{eq19}
H=-J\sum_{<i,j>}c_{i}c_{j}S_{i}^{z}S_{j}^{z}-D\sum_{i}c_{i}(S_{i}^{z})^{2}-\Omega\sum_{i}c_{i}S_{i}^{x},
\end{equation}
where $S_{i}^{z}$ and $S_{i}^{x}$ denote the $z$ and $x$ components of the spin operator, respectively. The first summation in Eq. (\ref{eq19}) is over the nearest-neighbor pairs of spins and the operator $S_{i}^{z}$ takes the values $S_{i}^{z}=0,\pm1$. $J$, $D$ and $\Omega$ terms stand for the exchange interaction, single-ion anisotropy (i.e. crystal field) and transverse field, respectively.

By using the approximated spin correlation identities \cite{sabarreto}
\begin{equation}\label{eq20}
c_{i}\left\langle\{f_{i}\}S_{i}^z\right\rangle=c_{i}\left\langle
\{f_{i}\} \frac{\mathrm{Tr}_{i}S_{i}^z\exp{(-\beta
H_{i})}}{\mathrm{Tr}_{i}\exp{(-\beta H_{i})}}\right\rangle,
\end{equation}
\begin{equation}\label{eq21}
c_{i}\left\langle\{f_{i}\}(S_{i}^{z})^2\right\rangle=c_{i}\left\langle
\{f_{i}\} \frac{\mathrm{Tr}_{i}(S_{i}^{z})^2\exp{(-\beta
H_{i})}}{\mathrm{Tr}_{i}\exp{(-\beta H_{i})}}\right\rangle,
\end{equation}
and following the same methodology of Sec. \ref{formualtion1}, we can obtain the general form of the site correlation functions for the central site as follows
\begin{equation}\label{eq22}
\left\langle c_{i}\left\langle \{f_{i}\} S_{i}^{z}\right\rangle\right\rangle_{r}=\left\langle c_{i}\left\langle\{f_{i}\}\prod_{j=1}^{q}\left[c_{j}\left(S_{j}^{z}\right)^{2}\cosh(J\nabla)
+c_{j}S_{j}^{z}\sinh(J\nabla)+1-c_{j}\left(S_{j}^{z}\right)^{2}\right]\right\rangle\right\rangle_{r}
F(x)|_{x=0},
\end{equation}
\begin{equation}\label{eq23}
\left\langle c_{i}\left\langle \{f_{i}\} (S_{i}^{z})^{2}\right\rangle\right\rangle_{r}=\left\langle c_{i}\left\langle\{f_{i}\}\prod_{j=1}^{q}\left[c_{j}\left(S_{j}^{z}\right)^{2}\cosh(J\nabla)
+c_{j}S_{j}^{z}\sinh(J\nabla)+1-c_{j}\left(S_{j}^{z}\right)^{2}\right]\right\rangle\right\rangle_{r}
G(x)|_{x=0},
\end{equation}
where the functions $F(x)$ ad $G(x)$ can be found in Ref. \cite{akýncý1}. By expanding the right hand side of Eqs. (\ref{eq22}) and (\ref{eq23}) according to Eq. (\ref{eq11}) for $c_{0}S_{0}^{z}$ and $c_{0}(S_{0}^{z})^{2}$, respectively with $\{f_{i}\}=1$ and taking only the nonzero terms we get magnetization and quadrupolar moment of the central site as follows
\begin{equation}\label{eq24}
m=\langle\langle c_{0}S_{0}\rangle\rangle_{r}=x_{1}=3cx_{4}k_{1}+cx_{6}k_{3}+(-6k_{1}+6k_{2})cx_{8}+(3k_{1}-6k_{2}+3k_{4})cx_{14},
\end{equation}
\begin{eqnarray}\label{eq25}
\nonumber
q_{z}=\langle\langle c_{0}S_{0}^{2}\rangle\rangle_{r}=x_{16}&=&cr_{0}+3cx_{5}r_{2}+(-3r_{0}+3r_{1})cx_{7}+(3r_{0}-6r_{1}+3r_{3})cx_{9}\\
&&+(-3r_{2}+3r_{4})cx_{13}+(-r_{0}+3r_{1}-3r_{3}+r_{5})cx_{15},
\end{eqnarray}
where the coefficients in Eqs. (\ref{eq24}) and (\ref{eq25}) are given as follows
\begin{eqnarray}\label{eq26}
\nonumber
&&r_{0}=G(0),\\
\nonumber
k_{1}=\sinh(J\nabla)F(x)|_{x=0},\ \ \ \ \ \ \ \ \ \ \ \ \ \ \ \ \ && r_{1}=\cosh(J\nabla)G(x)|_{x=0},\\
\nonumber
k_{2}=\cosh(J\nabla)\sinh(J\nabla)F(x)|_{x=0},\ \ \ \ && r_{2}=\sinh^{2}(J\nabla)G(x)|_{x=0},\\
\nonumber
k_{3}=\sinh^{3}(J\nabla)F(x)|_{x=0},\ \ \ \ \ \ \ \ \ \ \ \ \ \ \ \ && r_{3}=\cosh^{2}(J\nabla)G(x)|_{x=0},\\
\nonumber
k_{4}=\cosh^{2}(J\nabla)\sinh(J\nabla)F(x)|_{x=0},\ \ \ && r_{4}=\cosh(J\nabla)\sinh^{2}(J\nabla)G(x)|_{x=0},\\
&&r_{5}=\cosh^{3}(J\nabla)G(x)|_{x=0}.
\end{eqnarray}
If we use improved decoupling approximation given in Eq. (\ref{eq9}) then Eqs. (\ref{eq22}) and (\ref{eq23}) reduce to the following coupled equations
\begin{eqnarray}\label{eq27}
\nonumber
m&=&c\left[\cosh(J\nabla)+m\sinh(J\nabla)+1-q_{z}\right]^{3}F(x)|_{x=0},\\
&=&3c mk_{1}+cm^{3}k_{3}+(-6k_{1}+6k_{2})c m q_{z}+(3k_{1}-6k_{2}+3k_{4})cmq_{z}^{2},
\end{eqnarray}
\begin{eqnarray}\label{eq28}
\nonumber
q_{z}&=&c\left[q\cosh(J\nabla)+m\sinh(J\nabla)+1-q_{z}\right]^{3}G(x)|_{x=0},\\
\nonumber
&=&cr_{0}+3c m^{2}r_{2}+(-3r_{0}+3r_{1})c q_{z}+(3r_{0}-6r_{1}+3r_{3})cq_{z}^{2}\\
&&+(-3r_{2}+3r_{4})cm^{2}q_{z}+(-r_{0}+3r_{1}-3r_{3}+r_{5})cq_{z}^{3},
\end{eqnarray}
where $m=\langle\langle c_{i}S_{i}^{z}\rangle\rangle_{r}$ and $q_{z}=\langle\langle c_{i}(S_{i}^{z})^{2}\rangle\rangle_{r}$. Eqs. (\ref{eq27}) and (\ref{eq28}) are nothing but just the results obtained in Refs.\cite{tucker,saber1,tucker1,bobak4} which exposes the superiority of the present method.

Now, we should evaluate the thermal and configurational averages of the perimeter site correlations within the present formalism. Thus, corresponding to Eqs. (\ref{eq22}) and (\ref{eq23}) we have
\begin{equation}\label{eq29}
\left\langle c_{\delta}\left\langle \{f_{\delta}\}S_{\delta}^{z}\right\rangle\right\rangle_{r}=\left\langle c_{\delta}\left\langle \{f_{\delta}\}\left[c_{0}\left(S_{0}^{z}\right)^{2}\cosh(J\nabla)+c_{0}S_{0}^{z}\sinh(J\nabla)+1-c_{0}\left(S_{0}^{z}\right)^{2}\right]\right\rangle\right\rangle_{r} F(x+\gamma)|_{x=0},
\end{equation}
\begin{equation}\label{eq30}
\left\langle c_{\delta}\left\langle \{f_{\delta}\}(S_{\delta}^{z})^{2}\right\rangle\right\rangle_{r}=\left\langle c_{\delta}\left\langle \{f_{\delta}\}\left[c_{0}\left(S_{0}^{z}\right)^{2}\cosh(J\nabla)+c_{0}S_{0}^{z}\sinh(J\nabla)+1-c_{0}\left(S_{0}^{z}\right)^{2}\right]\right\rangle\right\rangle_{r} G(x+\gamma)|_{x=0},
\end{equation}
where $\gamma=(q-1)A$ represents the effective field produced by the $(q-1)$ spins outside of the cluster. From Eqs. (\ref{eq29}) and (\ref{eq30}) with $\delta=\{f_{\delta}\}=1$, we can get the perimeter site correlation functions as follows
\begin{equation}\label{eq31}
\langle\langle c_{1}S_{1}\rangle\rangle_{r}=x_{4}=a_{1}c+a_{2}cx_{1}+(a_{3}-a_{1})cx_{16},
\end{equation}
\begin{equation}\label{eq32}
\langle\langle c_{1}S_{1}^{2}\rangle\rangle_{r}=x_{7}=b_{1}c+b_{2}cx_{1}+(b_{3}-b_{1})cx_{16},
\end{equation}
where
\begin{eqnarray}\label{eq33}
\nonumber
a_{1}=F(\gamma),\ \ \ \ \ \ \ \ \ \ \ \ \ \ \ \ \ \ &&b_{1}=G(\gamma),\\
\nonumber
a_{2}=\sinh(J\nabla)F(x+\gamma),\ &&b_{2}=\sinh(J\nabla)G(x+\gamma),\\
a_{3}=\cosh(J\nabla)F(x+\gamma),\ &&b_{3}=\cosh(J\nabla)G(x+\gamma).
\end{eqnarray}
The coefficients in Eqs. (\ref{eq26}) and (\ref{eq33}) can be calculated by using the relation $e^{\alpha\nabla}f(x)=f(x+\alpha)$.

For a dilute spin-1 BC model, using Eqs. (\ref{eq22}), (\ref{eq23}), (\ref{eq29}) and (\ref{eq30}) we derive a set of linear equations by considering that (i) the correlations depend only on the distance between the lattice sites, (ii) the average values of a central site and its nearest-neighbor site (it is labeled as the perimeter site) are equal to each other with the fact that, in the matrix representations of spin operator $\hat{S}$, the spin-1 system has the properties $(S_{j}^{z})^{3}=S_{j}^{z}$ and $(S_{j}^{z})^{4}=(S_{j}^{z})^{2}$. Thus, the number of the set of linear equations obtained for the spin-1 Ising system with $q=3$ reduces to twenty one, and a detailed derivation and the complete set is given in \ref{appendixb}.

Since the thermal and configurational averages of the central site is equal to that of its nearest-neighbor sites within the present method then the unknown parameter $A$ in Eq. (\ref{eq33}) can be numerically determined by the relation
\begin{equation}\label{eq34}
x_{1}=x_{4}.
\end{equation}
By solving Eq. (\ref{eq34}) numerically at a given fixed set of Hamiltonian parameters we obtain the parameter $A$. Then we use the numerical values of $A$ to obtain the site correlation functions such as the longitudinal magnetization $\langle\langle c_{0}S_{0} \rangle\rangle_r$, longitudinal quadrupolar moment $\langle\langle c_{0}S_{0}^{2} \rangle\rangle_r$ and so on, which can be found from Eq. (\ref{eq2b}). $A=0$ always satisfies Eq. (\ref{eq34}) and gives paramagnetic solution. On the other hand, nonzero solutions of $A$ which satisfy Eq. (\ref{eq34}) just correspond to ferromagnetic state solutions of the system. The critical temperature $k_{B}T_{c}/J$ can be found by solving Eq. (\ref{eq34}) in the limit of $\gamma\rightarrow0$. Depending on the Hamiltonian parameters, there may be two solutions [i.e.,two critical temperature values satisfy Eq. (\ref{eq34})] corresponding to the first (or second) and second-order phase-transition points, respectively. We determine the type of the transition by looking at the temperature dependence of magnetization for selected values of system parameters.

\section{Results and discussion}\label{results}

\subsection{Site diluted spin-1/2 model}
\begin{figure}[!h]
\subfigure[\hspace{0 cm}] {\includegraphics[width=8cm]{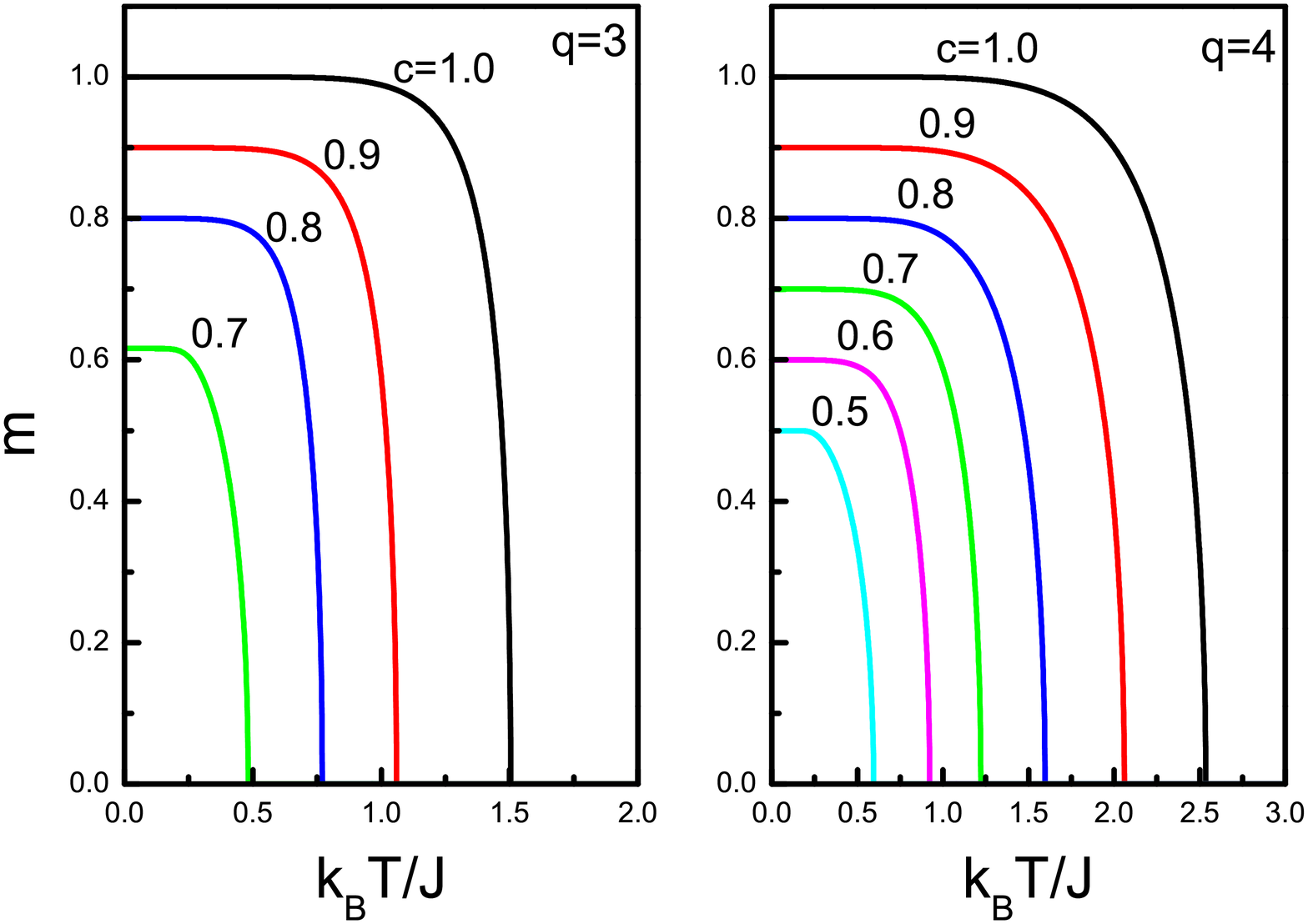}}
\subfigure[\hspace{0 cm}] {\includegraphics[width=8cm]{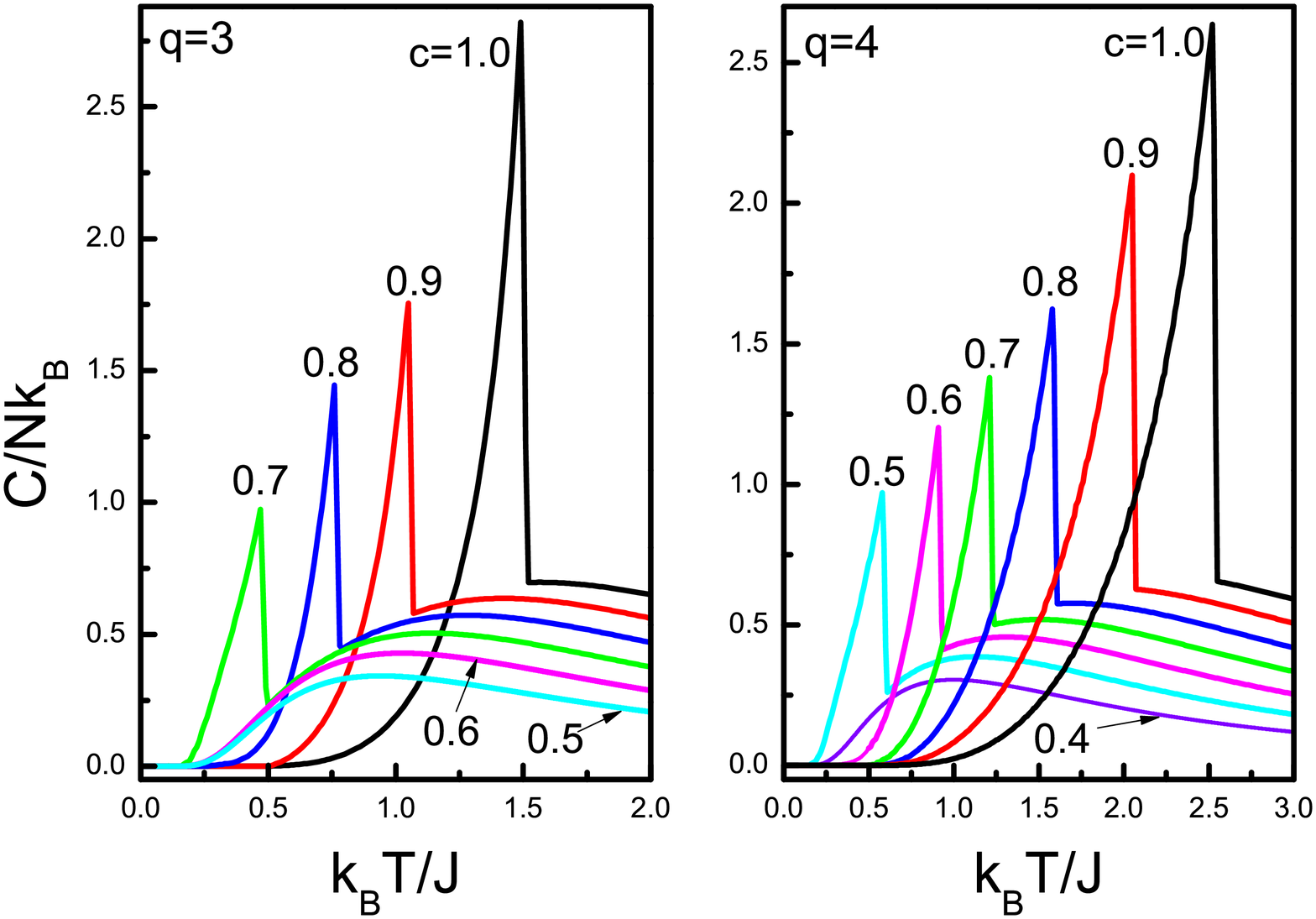}}\\
\subfigure[\hspace{0 cm}] {\includegraphics[width=8cm]{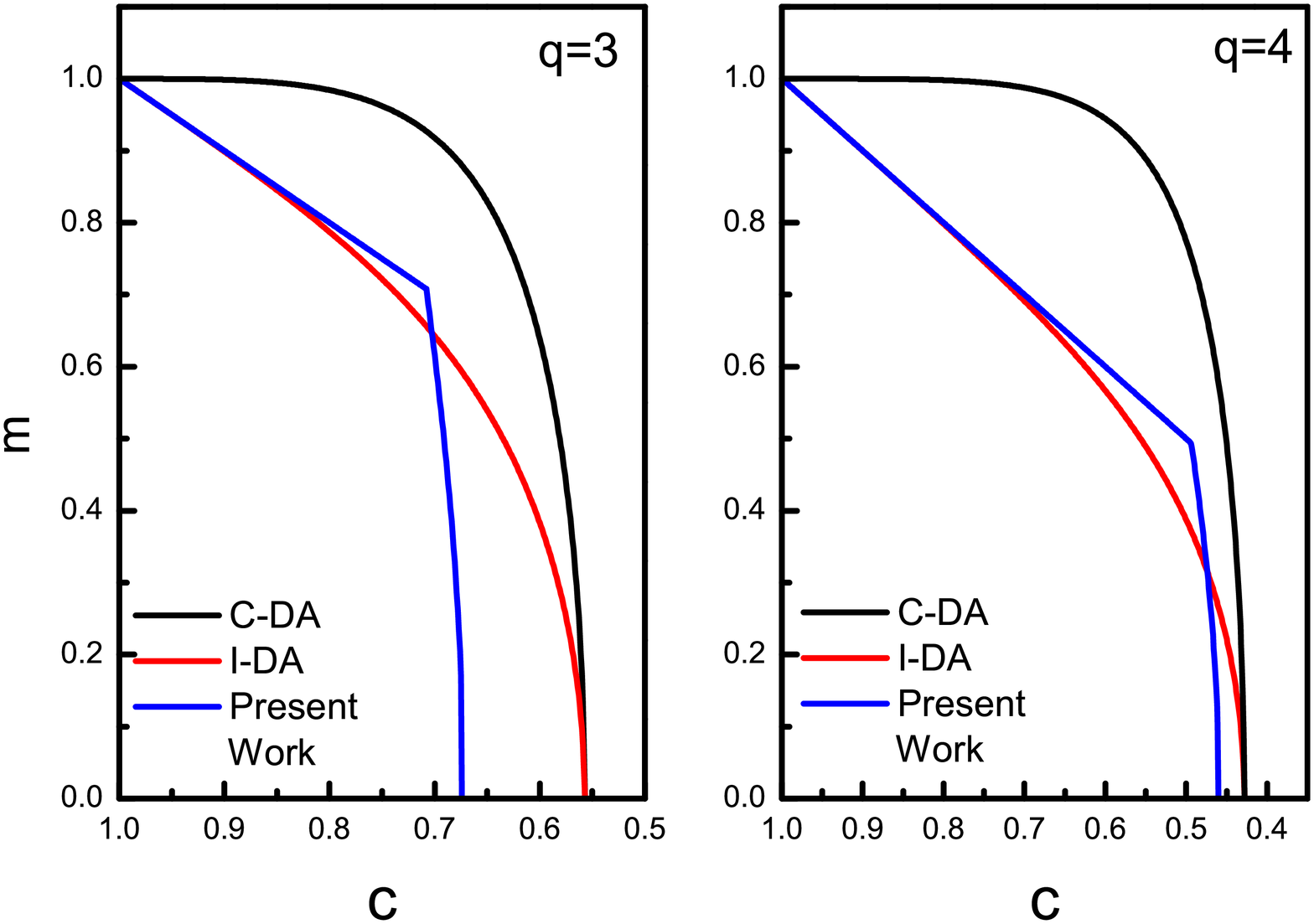}}
\subfigure[\hspace{0 cm}] {\includegraphics[width=8.25cm]{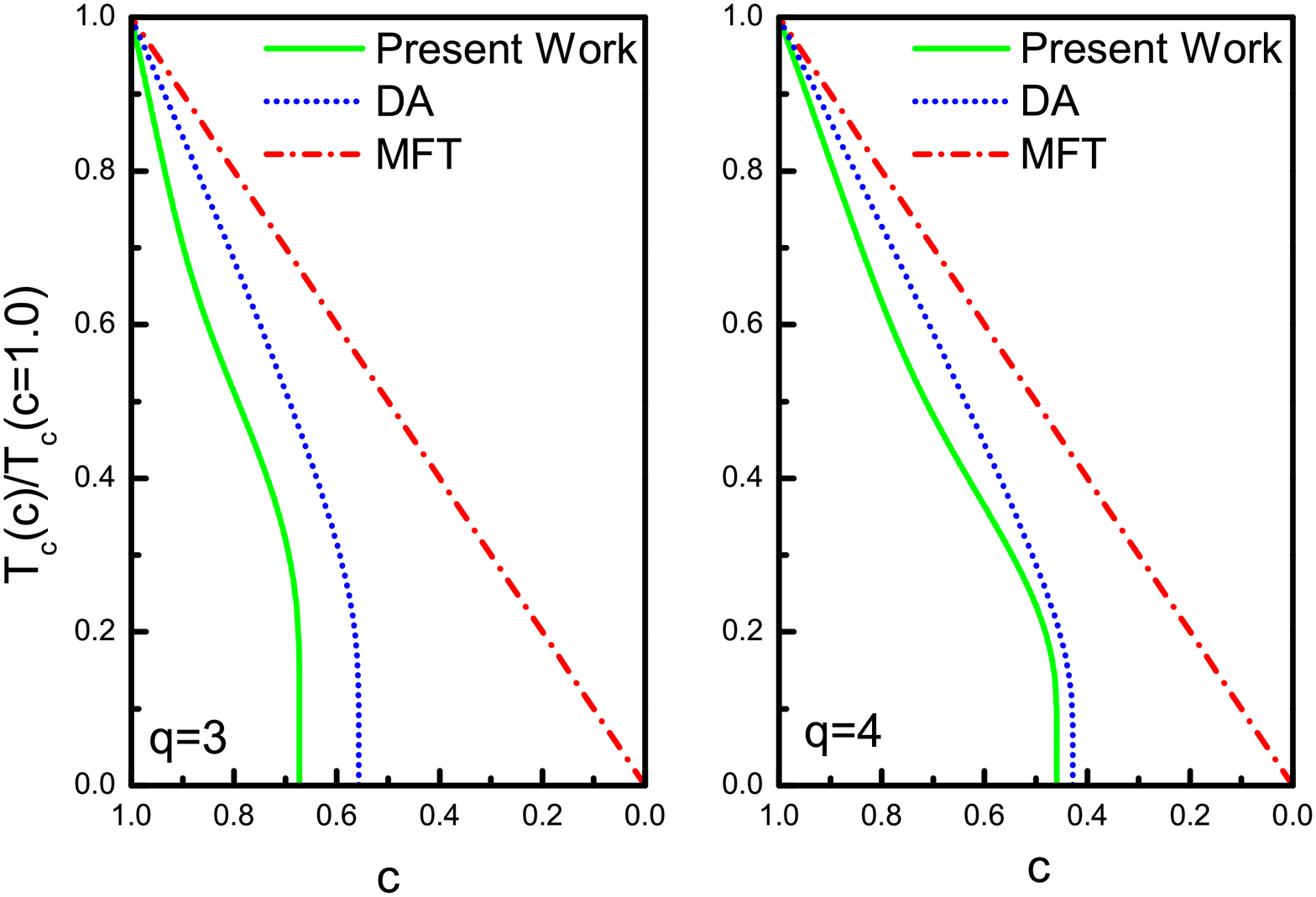}}\\
\caption{Temperature dependence of (a) magnetization, (b) specific heat curves of dilute ferromagnetic system for honeycomb $(q=3)$ and square $(q=4)$ lattices with some selected values of site concentration $c$. (c) Ground state magnetizations as a function of temperature for $q=3$, and $q=4$. (d) Phase diagrams of the system in $(k_{B}T_{c}/J-c)$ plane obtained by MFT (dash-dotted), DA (dotted), and present work (solid).}
\label{fig1}
\end{figure}

In Fig. (\ref{fig1}) we show the phase diagrams and magnetization, as well as specific heat curves for honeycomb $(q=3)$ and square $(q=4)$ lattices which can be obtained by solving Eqs. (\ref{eq2a}) and (\ref{eq4a}) numerically. In Fig. (\ref{fig1}a) variation of magnetization curves are depicted as a function of temperature $k_{B}T/J$ with typical values of site concentration $c$. As expected, we see in Fig. (\ref{fig1}a) that as the temperature increases starting from zero, the magnetization of the system decreases continuously, and it falls rapidly to zero at the critical temperature $k_{B}T_{c}/J$ for selected $c$ values. The number of interacting sites on the lattice decreases as $c$ decreases and hence, $k_{B}T_{c}/J$ value of the system and the saturation value of magnetization curves also decrease as $c$ decreases.

In Fig. (\ref{fig1}b) we examine the effect of site concentration $c$ on the temperature dependence of specific heat of the system. We see that as the temperature increases starting from zero, then the specific heat curves exhibit a sharp peak at a second-order phase transition temperature which decreases with decreasing $c$. As $c$ approaches its critical value $c^{*}$ at which critical temperature reduces to zero then an additional broad cusp appears and below $c^{*}$ phase transition disappears. For $c>c^{*}$ the system forms an infinite cluster of lattice sites however, as $c$ gets closer to $c^{*}$ then isolated finite clusters appear and for $c<c^{*}$ the system cannot exhibit long range ferromagnetic order even at zero temperature which causes a broad cusp in specific heat vs temperature curves. These observations are qualitatively agree with those of Refs. \cite{bobak1,Alcantara,balcerzak1,wiatrowski} and show the proper thermodynamic behavior over the whole range of temperatures, including the ground-state behavior $(C/Nk_{B}\rightarrow0 \quad\mathrm{as}\quad k_{B}T/J\rightarrow0)$ and the thermal stability condition $(C/Nk_{B}\geq0)$. Next, Fig. (\ref{fig1}c) represents the variation of the saturation magnetization with site concentration. In this figure, we also compare our results (blue line) with those of EFT based on conventional DA (C-DA, black line) and improved DA  (I-DA, red line) methods. It is clearly evident that site dilution lowers down the saturation magnetization. According to C-DA saturation magnetization of the system continuously decreases as $c$ decreases then falls rapidly to zero at $c^{*}$. On the other hand, I-DA predicts a linear decrease at high magnetic atom concentrations, but as $c$ decreases gradually then a monotonic decline is observed in the saturation magnetization value. On the other hand, according to our results we observe a linear decrement trend up to the vicinity of $c^{*}$ which originates as a result of considering the multi-site correlations. Finally, we represent the phase diagram of the system in $(k_{B}T_{c}/J-c)$ plane which separates the ferromagnetic and paramagnetic phases and we compare our results with those of the other methods in the literature. According to this figure, critical temperature $k_{B}T_{c}/J$ of system decreases gradually, and ferromagnetic region gets narrower as $c$ increases, and $k_{B}T_{c}/J$ value depresses to zero at $c=c^{*}$. Such a behavior is an expected fact in dilution problems. Numerical value of critical concentration $c^{*}$ for honeycomb $(q=3)$ and square $(q=4)$ lattices is given in Table \ref{table1}, and compared with the other works in the literature. It is well known that the series expansion (SE) method gives the best approximate values to the known exact results \cite{stauffer}. Therefore, we see in Table \ref{table1} that the present work improves the results of finite cluster approximation (OSCA and TSCA), as well as the other works based on EFT with DA. The reason is due to the fact that, in contrast to the previously published works mentioned above, there is no uncontrolled decoupling procedure used for the higher-order correlation functions within the present approximation.
\begin{table}[h!]
\small
\begin{center}
\begin{threeparttable}
\caption{Numerical values of critical site concentration $c^{*}$ for spin-1/2 system obtained within the present work for $q=3,4$ and comparison with various approximations in the literature: Average coordination number approximation $2/q$ and Bethe approximation $(q-1)^{-1}$ \cite{sato}, RG \cite{yeomans1,yeomans2}, CEFT \cite{taggart}, EFT \cite{balcerzak1,li,bobak2,saber3}, OSCA \cite{wiatrowski,bobak4}, TSCA \cite{bobak3,bobak4}, CVM \cite{balcerzak3}, MC \cite{neda}, SE \cite{sykes1,sykes2}.}
\renewcommand{\arraystretch}{1.2}
\begin{tabular}{lllllllllllll}
\thickhline
$q$ &MFT& $2/q$  & $(q-1)^{-1}$ & RG& CEFT & EFT &OSCA& TSCA & CVM &MC & SE & Present Work\\
\hline $3$  &0& 0.667 &  0.5 & &  0.711 &  0.5575  & 0.5575& 0.5706 &  0.768&   & 0.698 & \ 0.6727  \\
$4$   &0&0.5& 0.333 & 0.602 &   0.558 &  0.4284   & 0.4284  & 0.4303  & 0.640  & 0.413 & 0.593 & \ 0.4594 \\
\thickhline \\
\label{table1}
\end{tabular}
\end{threeparttable}
\end{center}
\end{table}
\begin{figure}[!h]
\subfigure[\hspace{0 cm}] {\includegraphics[width=8cm]{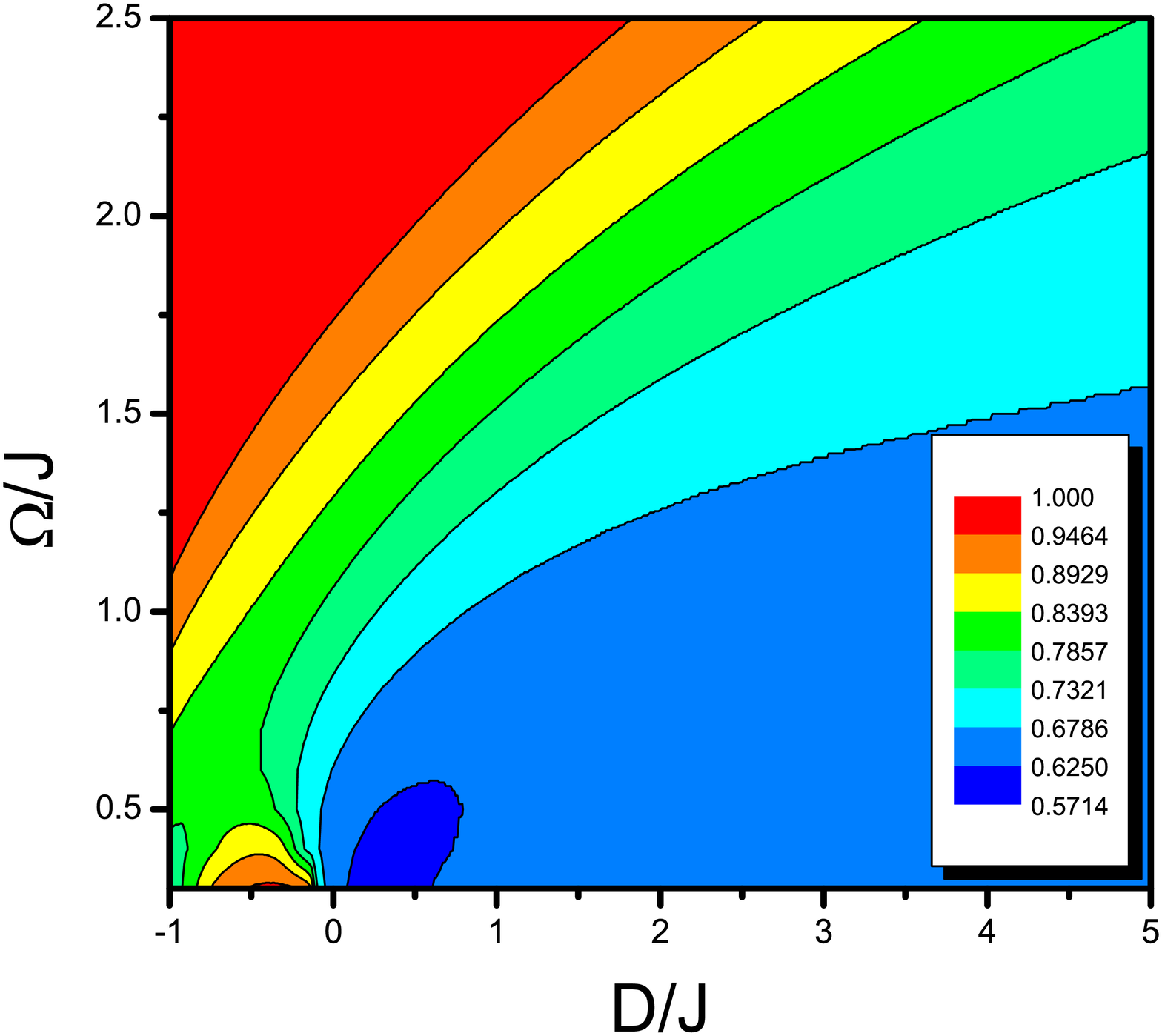}}
\subfigure[\hspace{0 cm}] {\includegraphics[width=8cm]{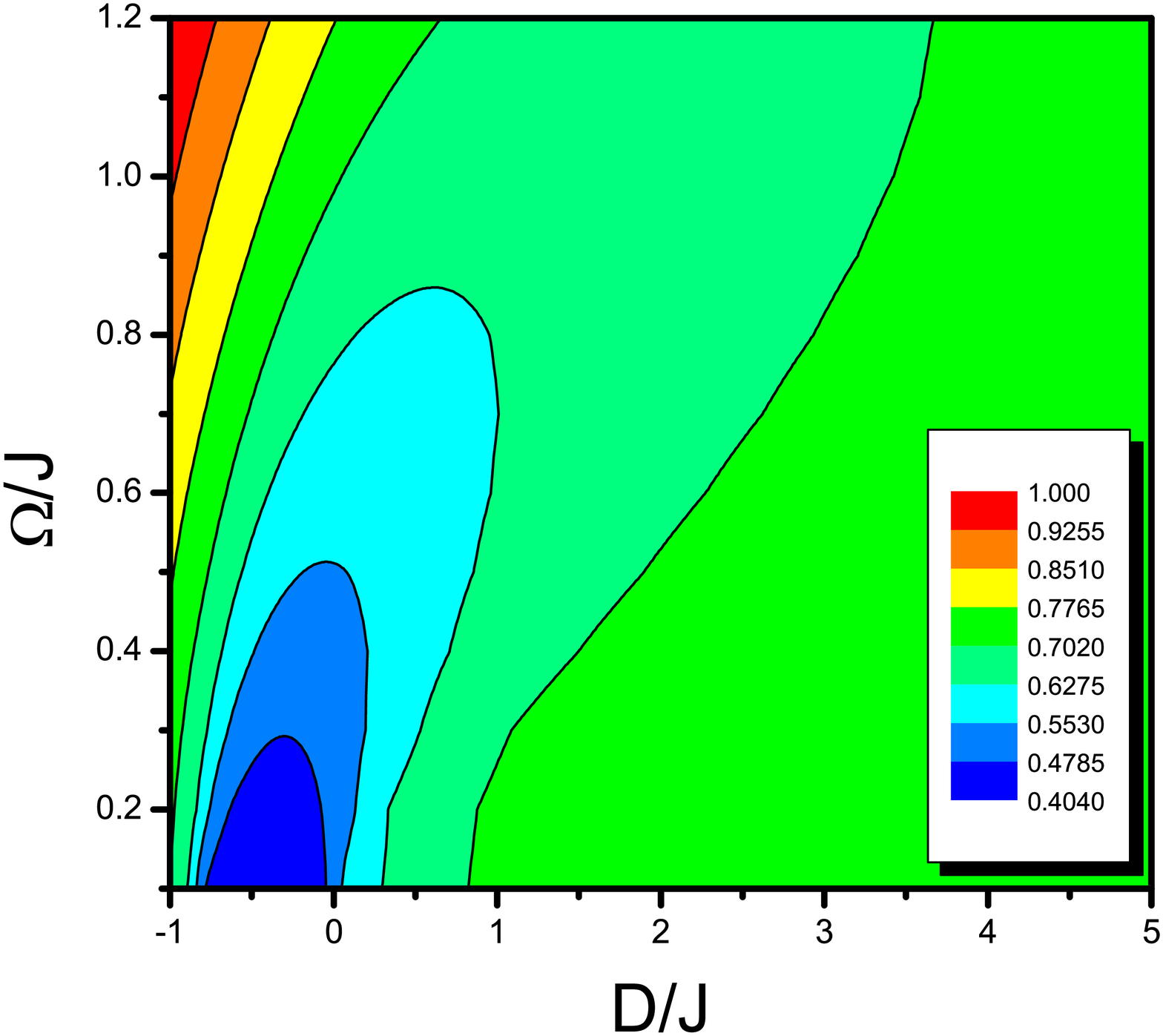}}\\
\caption{(a) Variation of critical site concentration of a diluted spin-1 BC model for $q=3$ with crystal field $D/J$ and transverse field $\Omega/J$. (b) Critical bond concentration of the same model projected on the same plane. (For interpretation of the references to color in these figure legends, the reader is referred to the web version of this article.)}
\label{fig2}
\end{figure}

\subsection{Site diluted spin-1 model}
For a site diluted spin-1 BC model defined by Hamiltonian (\ref{eq19}), we investigate the thermal and magnetic properties of the system by solving Eq. (\ref{eq2b}) numerically with condition (\ref{eq34}). At first, we shall examine the variation of the site percolation threshold $c^{*}$ with $D/J$ and $\Omega/J$. In Fig. (\ref{fig2}a) we plot the dependence of the site percolation threshold surface with $5.0<D/J<-1.0$ and $0.3<\Omega/J<2.5$. As we can see from Fig. (\ref{fig2}a), the effect of the transverse field $\Omega/J$ on the percolation threshold value clearly depends on the value of the crystal field $D/J$ and vice versa. Namely, for the values of $\Omega/J>0.565$ if we decrease the value of crystal field starting from $D/J=5.0$ then $c^{*}$ value increases and and reaches its maximum value. On the other hand, for $0.3<\Omega/J\leq0.565$ $c^{*}$ value increases or decreases depending on the value of $D/J$. Furthermore, for $\Omega/J\leq1.56$  and sufficiently large positive $D/J$, $c^{*}$ value remains more or less constant and we obtain $c^{*}=0.6727$ which is the critical site concentration of spin-1/2 system for $q=3$. Besides, for $D/J=0$ and $\Omega/J=0$ we get $c^{*}=0.6211$ which is higher than the bond percolation threshold value of the same system obtained by the same method \cite{akýncý1}. This value can be compared with the results obtained by the other works given in Table \ref{table2}. In Table \ref{table2}, two different critical concentrations obtained by EFT comes from the usage of exact or approximate Van der Waerden identity. Using the exact identity one obtains the result of OSCA. By comparing Table \ref{table1} and Table \ref{table2} we see that critical site concentration $c^{*}$ of a dilute system depends on the spin value $S$. However, according to the percolation theory \cite{ziman,stauffer} $c^{*}$ only depends on the topology of the lattice and must be independent of $S$. In order to fix this problem, Refs. \cite{kaneyoshi4,kaneyoshi6,kaneyoshi7} suggested to include a positive crystal field $D/J$ but, it is clear in Fig. (\ref{fig2}) that there is an exceptional situation (dark blue region in Fig. (\ref{fig2}b)) due to the presence of $\Omega/J$. Therefore, we can say that topology deformation of the percolation threshold surface illustrated in Fig. (\ref{fig2}) originates from a competition due to the presence of $D/J$ and $\Omega/J$ in the system. For completeness of the work, we also give the critical bond concentration surface of the same model obtained by the same methodology presented in this paper for $q=3$ \cite{akýncý1}. By drawing inspiration from Figs. (\ref{fig2}a) and (\ref{fig2}b), we think that whether in a site or bond dilution problem, the mechanism underlying the complex topological behavior of the critical concentration completely originates from a collective effect of both $\Omega/J$ and $D/J$.
\begin{figure}
\center
\includegraphics[width=12cm]{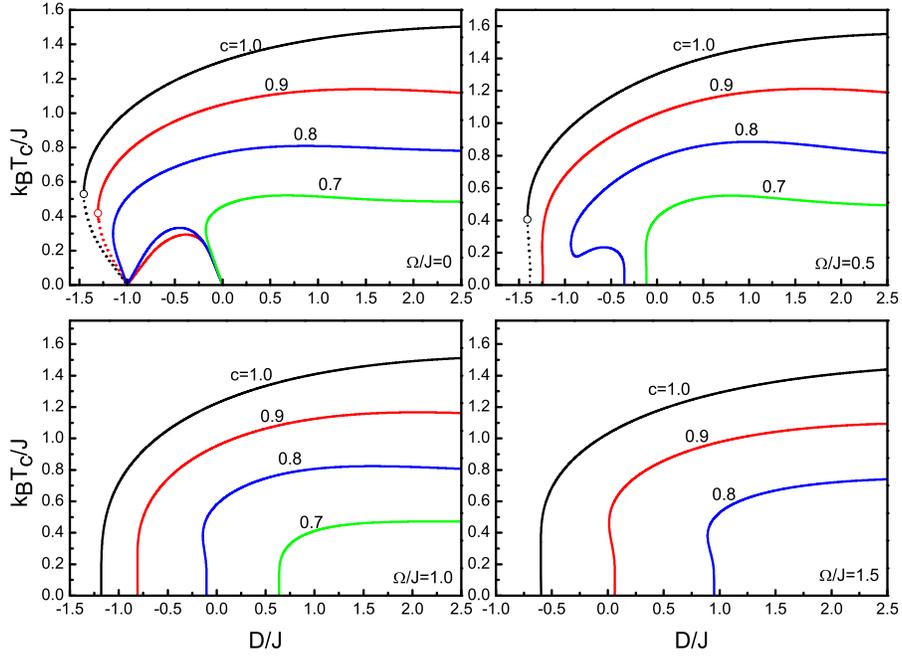}\\
\caption{The phase diagrams of a diluted spin-1 BC model in $(k_{B}T_{c}/J-D/J)$ plane for $q=3$ with selected values of transverse field $\Omega/J=0.0, 0.5, 1.0$ and $1.5$. Solid and dotted curves correspond to the second and first order phase transitions, respectively. Solid circles represent the tricritical points, and the numbers on the curves denote the site concentration $c$.}
\label{fig3}
\end{figure}
\begin{table}[h]
\begin{center}
\begin{threeparttable}
\caption{Site percolation threshold value $c^{*}$ for $D/J=0$ and  $\Omega/J=0$ obtained by present work for spin-1 system on a honeycomb lattice. For comparison, the results obtained by OSCA and TSCA \cite{bobak4}, EFT \cite{kaneyoshi6,kaneyoshi7} and SE \cite{sykes1,sykes2} are also given.}
\renewcommand{\arraystretch}{1.3}
\begin{tabular}{llllllll}
\thickhline
OSCA & TSCA & EFT & EFT  & SE &  Present Work\\
\hline  0.5158 &  0.5449& 0.5085&  0.5158&  0.698&  0.6211 \\
\thickhline \\
\label{table2}
\end{tabular}
\end{threeparttable}
\end{center}
\end{table}

In Fig. (\ref{fig3}), we represent the phase diagrams of the system in $(k_{B}T_{c}/J-D/J)$ plane for $\Omega/J=0$, $0.5$, $1.0$ and $1.5$ where the solid and dotted lines correspond to the second and first order transitions and hollow circles denote the tricritical points. The numbers accompanying each curve denote the value of site concentration $c$. In Fig. (\ref{fig3}), it is obvious that diluting the lattice sites reduces the critical temperatures of the second order phase transitions in the system for $D/J\geq0$. As seen in the upper left panel of Fig. (\ref{fig3}), the curve corresponding to pure case ($\Omega/J=0$ and $c=1.0$) exhibits a reentrant behavior of first order where a second order phase transition is followed by a first order phase transition at low temperatures for certain negative values of $D/J$. On the other hand, for $c=0.9$ we observe an extraordinary feature in the phase diagrams. In other words, there are two regions in $D/J$ plane at which a reentrant behavior occurs. The usual one is located within the interval $-1.3059<D/J<-1.0$ with a tricritical point $(D_{t}/J=-1.3058, k_{B}T_{t}/J=0.4192)$, and the other is found between $-1.0<D/J<0.0$. The latter behavior is quite interesting, since another tricritical point appears at $D_{t}/J=-1.0\quad \mathrm{and}\quad k_{B}T_{t}/J=0.0$. Besides, for $0>D/J>-1.0$ the system exhibits a reentrant behavior of second order. On the other hand, if we select $c=0.8$ then we see that the first order phase transitions and tricritical points disappear, and the system exhibits a reentrant behavior of second order within the interval $-1.1471<D/J<0.0$. Furthermore, the reentrance disappears as $D/J$ becomes positive for all selected values of $c$. Meanwhile, $(k_{B}T_{c}/J-D/J)$ phase diagrams for some selected values of $c$ with $\Omega/J=0.5$ are depicted on the upper right panel in Fig. (\ref{fig3}). It is clearly seen from this figure that the system exhibits a first order reentrance for $c=1.0$ only in a narrow region $-1.4077<D/J<-1.3796$. For $c=0.9$, tricritical point and reentrance tends to disappear, but if we decrease the magnetic atom concentration further, such as for $c=0.8$ then the phase diagrams exhibit a bulge with a pronounced second order reentrance within the interval $-0.9358<D/J<-0.3554$. If we select sufficiently large transverse field strengths, such as $\Omega/J=1.0$, and $1.5$ then the system cannot exhibit first order transitions and tricritical points anymore, even if $c=1.0$. In this case, we observe only second order phase transitions and ferromagnetic region gets narrower as $c$ decreases. In Ref.\cite{htoutou1}, the authors studied the same model for $q=4$, but they have not reported the behavior shown in Fig.(\ref{fig3}) in their paper. All of the observations reported here can also be verified by examining the corresponding magnetization curves (see Fig.\ref{fig7}).
\begin{figure}[!h]
\subfigure[\hspace{0 cm}] {\includegraphics[width=8cm]{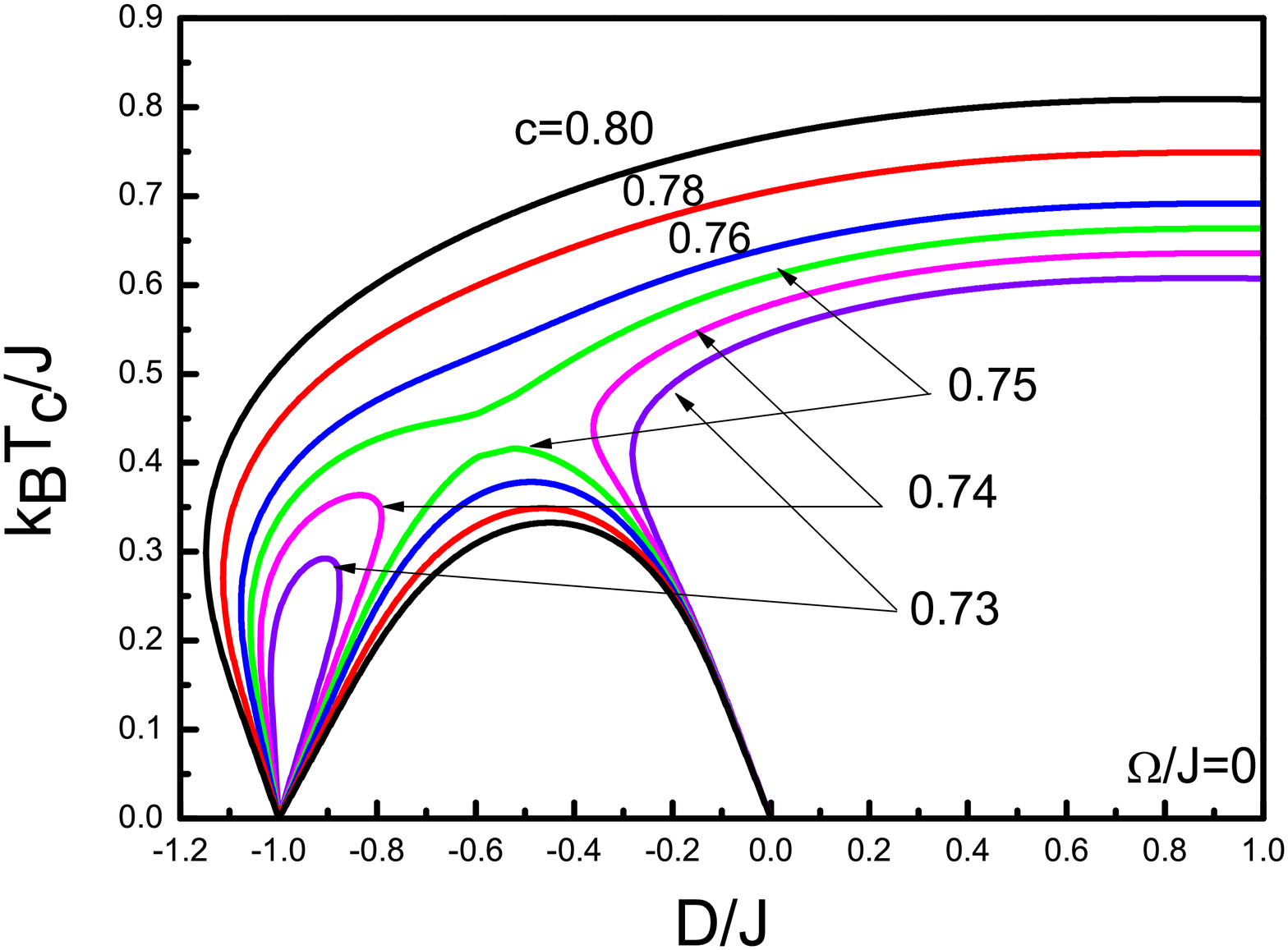}}
\subfigure[\hspace{0 cm}] {\includegraphics[width=8cm]{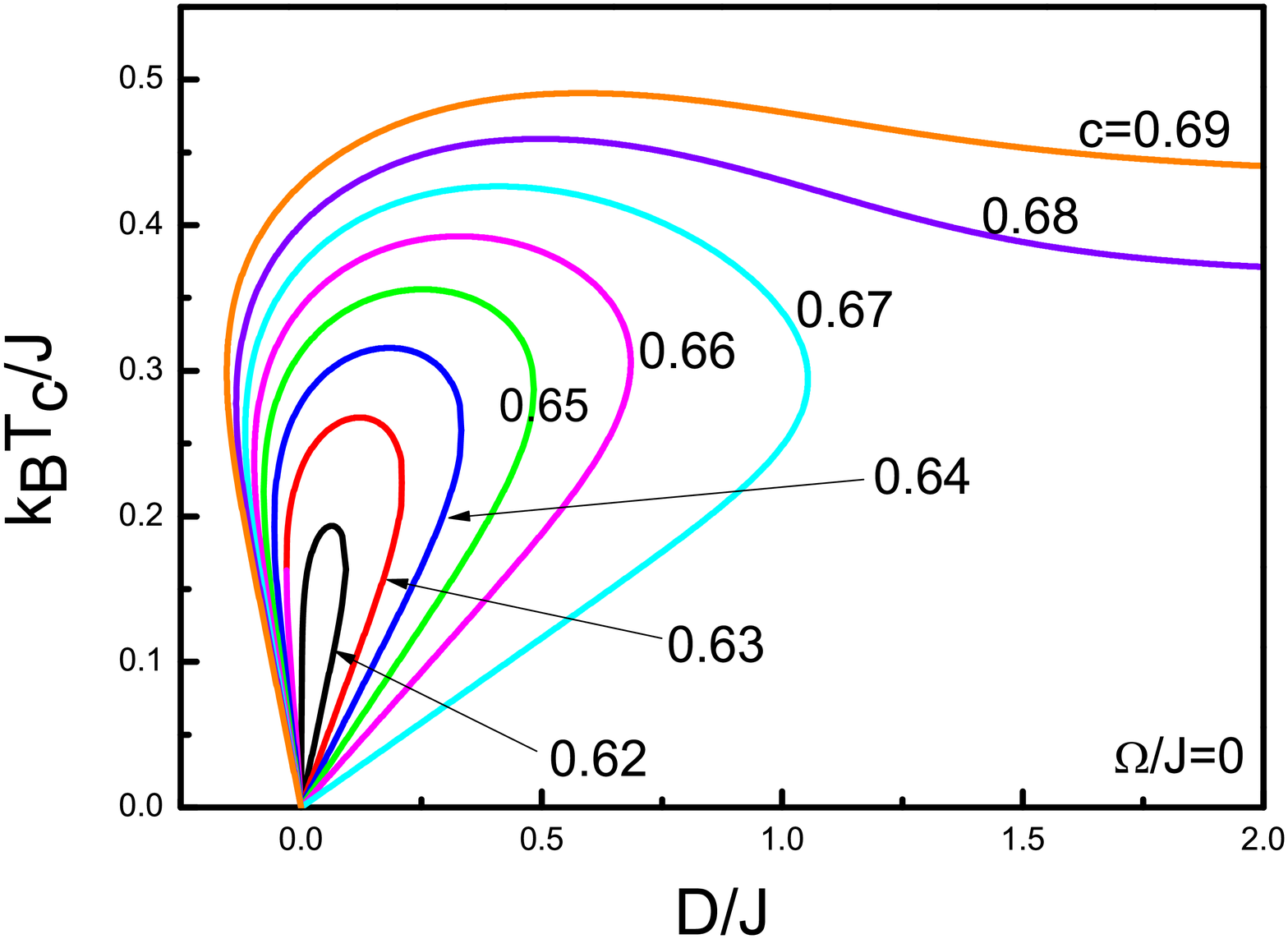}}\\
\subfigure[\hspace{0 cm}] {\includegraphics[width=8cm]{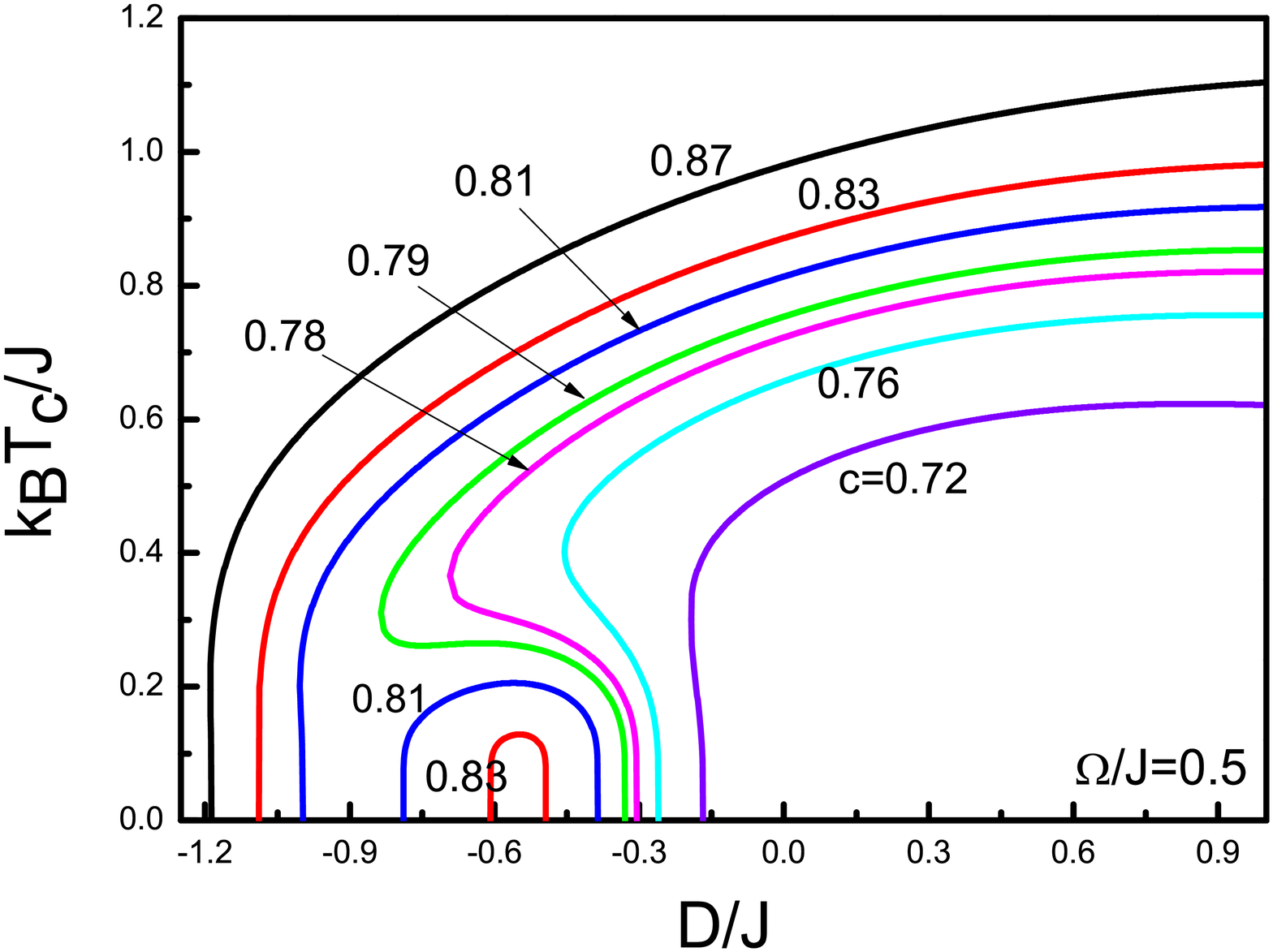}}
\subfigure[\hspace{0 cm}] {\includegraphics[width=8cm]{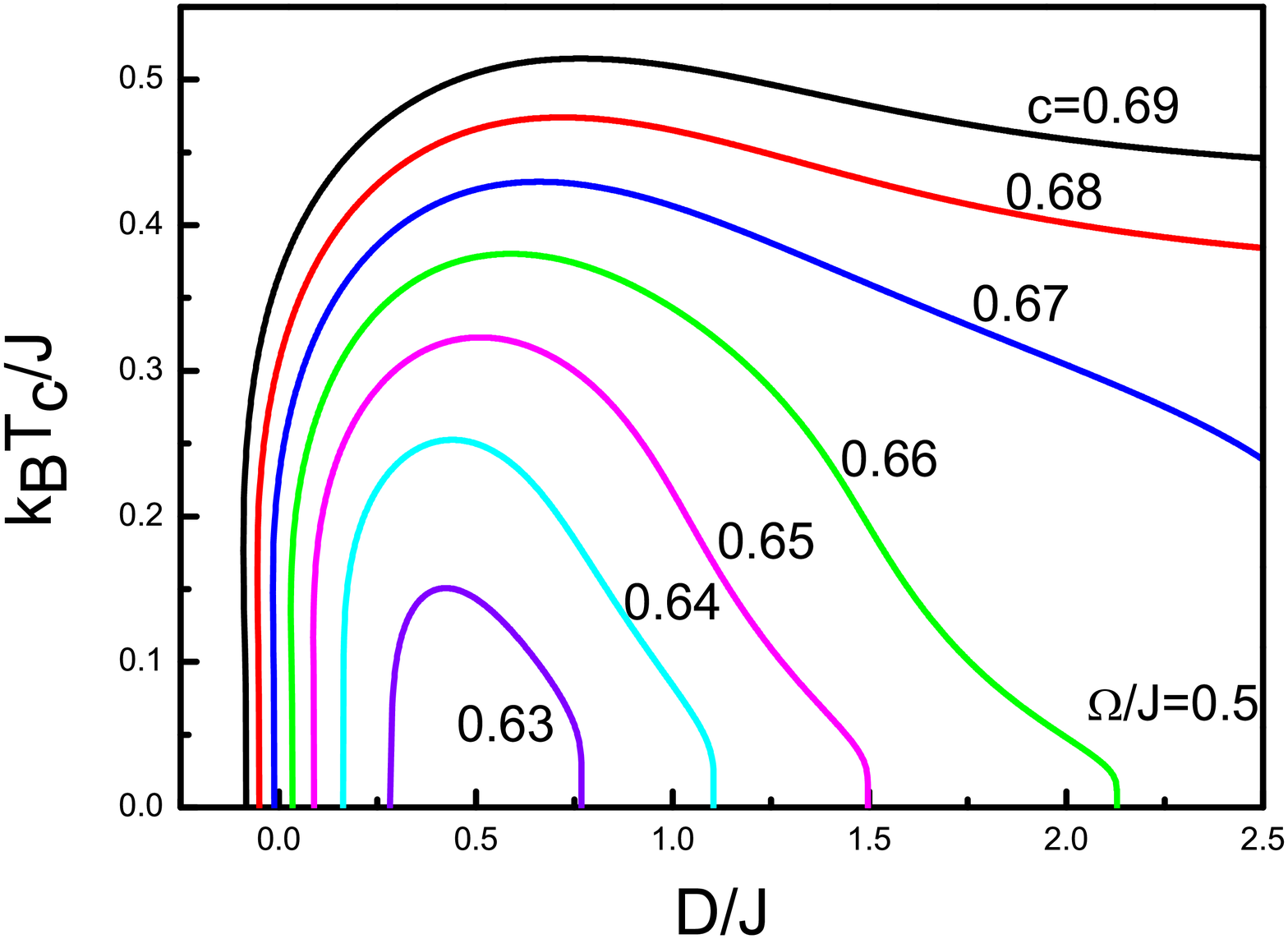}}\\
\caption{Evolution of the phase diagrams given in Fig.  (\ref{fig3}) for (a) $\Omega/J=0.0$ and $0.73\leq c \leq0.80$, (b) $\Omega/J=0.0$ and $0.62\leq c \leq0.69$, (c) $\Omega/J=0.5$ and $0.72\leq c \leq0.87$, (d) $\Omega/J=0.5$ and $0.63\leq c \leq0.69$. The numbers accompanying each curve denote the site concentration $c$.}
\label{fig4}
\end{figure}

\begin{figure}
\center
\includegraphics[width=12cm]{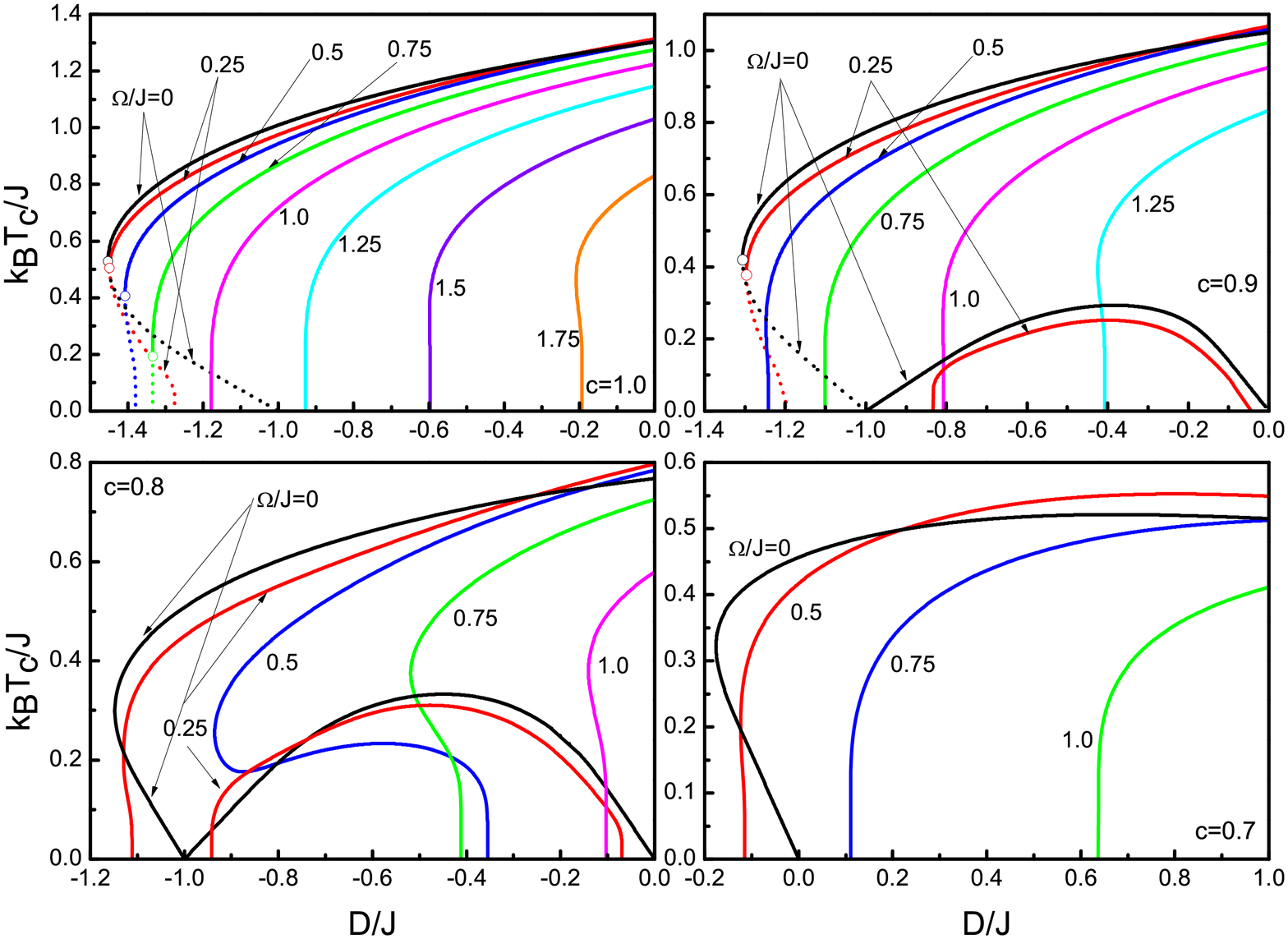}\\
\caption{The phase diagrams of a diluted spin-1 BC model in $(k_{B}T_{c}/J-D/J)$ plane for $q=3$ with selected values of site concentration $c=1.0, 0.9, 0.8$ and $0.7$. Solid and dotted curves correspond to the second and first order phase transitions, respectively. Solid circles represent the tricritical points, and the numbers on the curves denote the value of transverse field $\Omega/J$.}
\label{fig5}
\end{figure}

The evolution of the phase diagrams shown in Fig. (\ref{fig3}) for $\Omega/J=0$ and $0.5$ are depicted in Fig. (\ref{fig4}). As seen in Fig. (\ref{fig4}a) where $\Omega/J=0$, the phase diagrams exhibit a reentrant behavior of second order within the interval $-1.0<D/J<0.0$ for $c=0.80$, $0.78$, $0.76$ and $0.75$ whereas the reentrant phase transition region for $c=0.74$ and $0.73$ is divided into two parts: The first part is located in the vicinity of $D/J=-1.0$ which gets narrower as $c$ decreases while the second one is observed between $-0.3607<D/J<0.0$. If we decrease $c$ further, such as for $c=0.69$ and $0.68$ (see Fig. (\ref{fig4}b)) reentrance disappears. However, for $0.62\leq c\leq0.67$ another reentrant regime appears, but now for $D/J\geq0$ which gets narrower as $c$ decreases.
Similarly, Figs. (\ref{fig4}c) and (\ref{fig4}d) represents the evolution of phase diagrams of the system corresponding to the upper right panel in Fig. (\ref{fig3}) where $\Omega/J=0.5$. As seen in Fig. (\ref{fig4}c), the system undergoes only a second order phase transition for $c=0.87$. However for lower site concentrations e.g. $c=0.83$ and $0.81$, reentrant behavior of second order appears between $-0.609<D/J<-0.4938$ and $-0.7891<D/J<-0.386$, respectively. In addition, for $c\leq0.79$ the system exhibits a bulge in the reentrant phase transition regime which gets narrower as $c$ decreases. For $c<0.72$ (Fig.(\ref{fig4}d)),  reentrance disappears, and the system undergoes a second order phase transition form a paramagnetic state to a ferromagnetic state with increasing temperature. For $c<0.63$, site concentration of the system approaches to percolation threshold $c^{*}$, hence the ferromagnetic region becomes fairly narrow.

Next, as a complementary investigation of Fig. (\ref{fig3}), we investigate the effects of transverse field interactions $\Omega/J$ on the phase diagrams of the system in Fig. (\ref{fig5}) for some selected values of $c$. The upper left panel in Fig. (\ref{fig5}) corresponds to the phase diagrams of the pure system \cite{yuksel1} where we see that reentrant phase transitions tend to disappear and tricritical points decrease as $\Omega/J$ decreases for $c=1.0$. Moreover, according to the upper right panel in Fig. (\ref{fig5}), another ferromagnetic phase boundary arises between $-1.0\leq D/J\leq0.0$ and $-0.833\leq D/J\leq-0.0434$ for $c=0.9$ and $\Omega/J=0.0, 0.25$, respectively. These additional phase transition lines get narrower as $\Omega/J$ increases, and disappear after a certain value of $\Omega/J$. Furthermore, phase diagrams shown in lower left and right panels invariably exhibit second order phase transitions for $c=0.8$ and $0.7$ which is independent from transverse field value, and the reentrance is not observed anymore for $c=0.7$ and $\Omega/J\geq0.5$.

\begin{figure}[!h]
\center
\includegraphics[width=12cm]{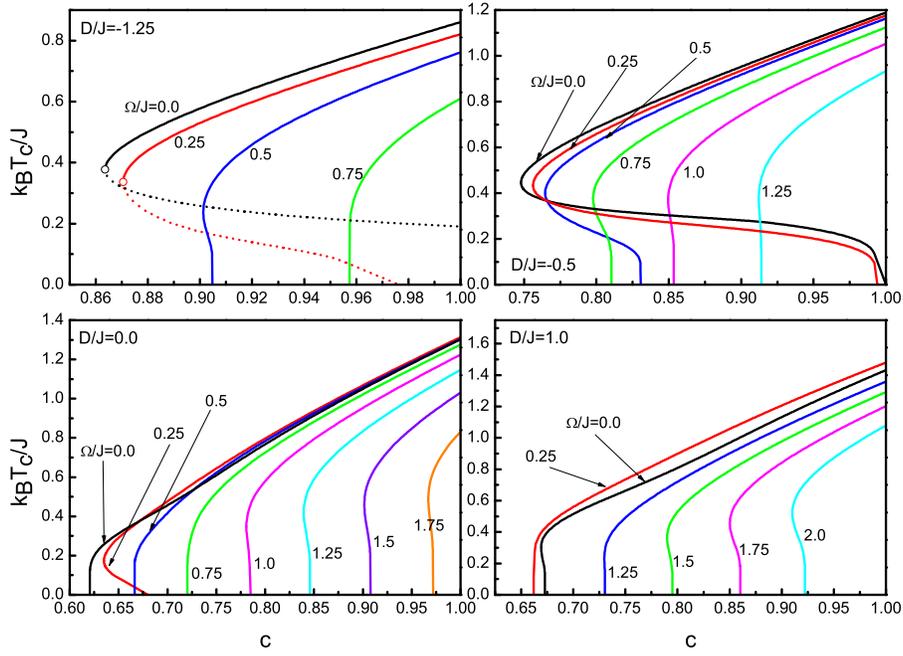}\\
\caption{The phase diagrams of a diluted spin-1 BC model in $(k_{B}T_{c}/J-c)$ plane for $q=3$ with selected values of the crystal field $D/J=-1.25, -0.5, 0.0$ and $0.5$. Solid and dotted curves correspond to the second and first order phase transitions, respectively. Solid circle denotes the tricritical point, and each curve in panels is plotted for a specific transverse field value.}
\label{fig6}
\end{figure}

\begin{figure}[!h]
\center
\includegraphics[width=12cm]{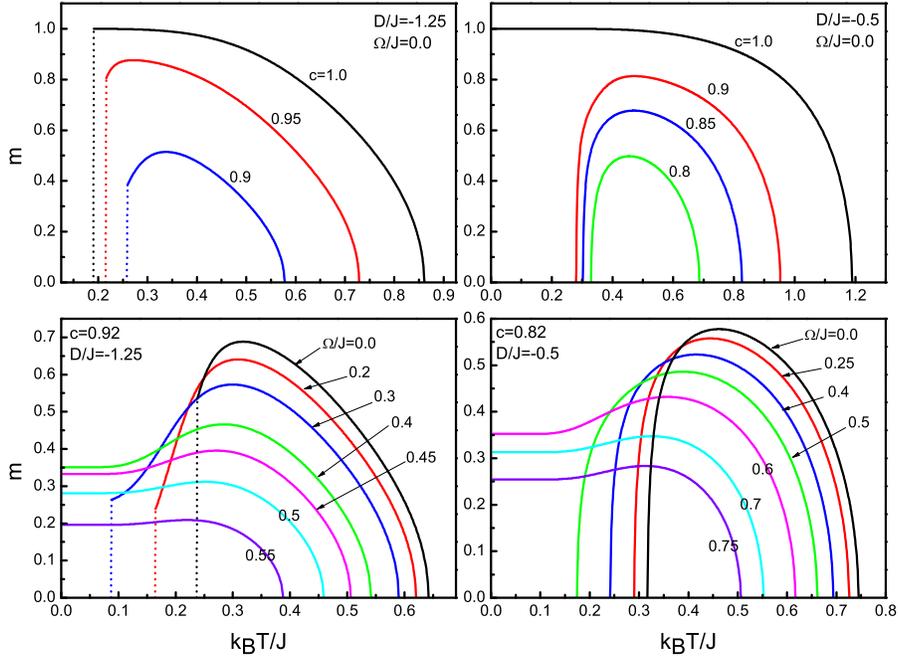}\\
\caption{Temperature dependence of the magnetization curves as functions of Hamiltonian parameters $D/J$, $\Omega/J$ and $c$. Solid and dotted magnetization curves exhibit second and first order phase transition properties, respectively.}
\label{fig7}
\end{figure}

In Fig. (\ref{fig6}), we examine the phase diagrams of the system in a $(k_{B}T_{c}/J-c)$ plane for $D/J=-1.25, -0.5, 0.0$ and $1.0$ with some typical $\Omega/J$ values. As seen in this figure, the system exhibits a tricritical behavior and its critical temperature cannot reach zero for $D/J=-1.25$ and $\Omega/J=0.0$, hence we cannot speak on any percolation threshold value. However, phase transition temperature of the system reduces to zero at $c^{*}=0.9764$ for $\Omega/J=0.25$. It is clear from upper left panel in Fig. (\ref{fig6}) that $c^{*}$ value for $\Omega/J=0.75$ is greater than those of $\Omega/J=0.5$, however it is lower than those of $\Omega/J=0.25$. On the other hand, for $D/J=-1.25$ and $\Omega/J=0.5$ the first order phase transitions and tricritical points disappear, and we barely observe a second order reentrance which also disappears for $\Omega/J=0.75$. If we select $D/J=-0.5$ then we cannot see any evidence of first order phase transitions, and critical site concentration $c^{*}$ of the system decreases as $\Omega/J$ increases up to $\Omega/J=0.75$. For $\Omega/J\geq 0.75$ we observe that $c^{*}$ value tends to increase as $\Omega/J$ increases which is consistent with the indications depicted in Fig. (\ref{fig2}a). Similar discussions are also valid for $D/J=0.0$ and $1.0$. Additionally, as an interesting characteristic of the system, we may note that the conditions for the occurrence of a second order reentrance in the system is rather complicated, since the existence or extinction of reentrance is rather sensitive to the collective effects of $D/J$, $\Omega/J$ and $c$.

As a final investigation, let us represent the temperature dependence of the magnetization for some selected values of Hamiltonian parameters corresponding to the phase diagrams depicted throughout Figs. (\ref{fig3})-(\ref{fig6}). In Fig. (\ref{fig7}), the typical transition profiles are shown for several values of Hamiltonian parameters. For example, if we select $D/J=-1.25$ and $\Omega/J=0.0$  magnetization curves corresponding to $c=1.0$, $0.95$ and $0.90$ exhibit a discontinuous jump at a first order transition temperature then gradually decrease and reduce to zero at a second order phase transition temperature with increasing temperature. This is an example of reentrance of first order in which a first order transition is followed by a second order transition. As an example of second order reentrant behavior, we can take a look at the magnetization curves that exist in a second order reentrant regime. For instance, we see that the magnetization curves exhibit two critical temperatures of the second order for $c=0.90$, $0.85$ and $0.80$ with $D/J=-0.5$ and $\Omega/J=0.0$. On the other hand, as an example of a second order ferromagnetic-paramagnetic phase transition, magnetization curves exhibit two different characteristics. As an example of the first case, we see that as the temperature increases then the magnetization of the system falls gradually from its saturation magnetization value at $k_{B}T/J=0.0$ and decreases continuously up to the vicinity of the transition temperature and vanishes at a critical temperature $k_{B}T_{c}/J$ for $c=1.0$ with $D/J=-0.5$ and $\Omega/J=0.0$ (corresponding to pure case), whereas in the second case the magnetization of the system exhibits a temperature-induced maximum with increasing temperature which is depicted on the lower left and right panels of Fig. (\ref{fig7}).

\section{Concluding remarks}\label{conclude}
In this work, we have investigated the thermal and magnetic properties of a site diluted spin-1/2 Ising model and a spin-1 Blume Capel (BC) model in the presence of transverse field interactions. We have introduced an effective-field approximation that takes into account the multi-site correlations in the cluster of a considered lattice with an improved configurational averaging technique. Our method is capable of locating the possible first order phase transition temperatures, as well as tricritical points, and under certain simplifications, equations of state obtained within the present approximation can be reduced to those obtained by conventional or improved decoupling approximation techniques which exposes the superiority of the present work.

For a spin-1/2 Ising system, we have obtained results that are superior to those estimated by conventional mean field theory (MFT) and effective field theory (EFT) based on a decoupling approximation, especially for the critical site concentration (i.e. site percolation threshold) value $c^{*}$ for honeycomb $(q=3)$ and square $(q=3)$ lattices. Our estimated values $c^{*}=0.6727$ and $c^{*}=0.4594$ for $q=3$ and $q=4$, respectively are the best approximate values to the results of MC and SE methods among the other works based on MFT or EFT.

In particular, we have investigated the phase diagrams and magnetization curves of a site diluted spin-1 BC model in the presence of transverse field interactions and we have shown that diluting the lattice sites may cause some drastic changes on some of the characteristic features of the model. For this model, we have examined the variation of the site percolation threshold $c^{*}$ with the crystal and transverse field interactions which has not been reported in the literature before. In the absence of crystal and transverse fields, the percolation threshold value of a site diluted spin-1 model for $q=3$ is estimated as $c^{*}$=0.6211 which improves the results obtained by other EFT based approximations. In addition, we have found that the percolation threshold value $c^{*}$ strictly depends on the value of crystal and transverse field interactions, as well as the topology of the lattice. We have also given the global phase diagrams, especially the first order phase transition lines that include reentrant phase transition regions. The results presented in this paper clearly indicate that the conditions for the occurrence of a second order reentrance in the system is rather complicated, since the existence or extinction of reentrance is rather sensitive to the competing effects between $D/J$, $\Omega/J$ and $c$. These observations cannot be observed by ignoring any of these Hamiltonian parameters in the system.

As a result, we can conclude that all of the points mentioned above show that our method improves the conventional EFT methods based on decoupling approximation. Therefore, we hope that the results
obtained in this work may be beneficial from both theoretical and experimental points of view.

\section*{Acknowledgements}
One of the authors (Y.Y.) would like to thank the Scientific and Technological Research Council of Turkey (T\"{U}B\.{I}TAK) for partial financial support. This work has been completed at Dokuz Eyl\"{u}l University, Graduate School of Natural and Applied Sciences, and the numerical calculations reported in this paper were performed at T\"{U}B\.{I}TAK ULAKBIM, High Performance and Grid Computing Center (TR-Grid e-Infrastructure). Partial financial support from SRF (Scientific Research Fund) of Dokuz Eyl\"{u}l University (2009.KB.FEN.077) (H.P.) is also acknowledged.

\newpage
\appendix
\section{Derivation of complete set of linear equations for spin-1/2 Ising Model}\label{appendixa}
In the present formalism, all of the site correlations including central, as well as perimeter site magnetizations are denoted by $x_{i}$. For instance, for a honeycomb lattice $(q=3)$ we have
\begin{eqnarray}\label{eq1a}
\nonumber
x_{1}&=&\langle\langle c_{0}S_{0}\rangle\rangle_{r},\\
\nonumber
x_{2}&=&\langle\langle c_{0}S_{0}c_{1}S_{1}\rangle\rangle_{r},\\
\nonumber
x_{3}&=&\langle\langle c_{0}S_{0}c_{1}S_{1}c_{2}S_{2}\rangle\rangle_{r},\\
\nonumber
x_{4}&=&\langle\langle c_{1}S_{1}\rangle\rangle_{r},\\
\nonumber
x_{5}&=&\langle\langle c_{1}S_{1}c_{2}S_{2}\rangle\rangle_{r},\\
x_{6}&=&\langle\langle c_{1}S_{1}c_{2}S_{2}c_{3}S_{3}\rangle\rangle_{r}.
\end{eqnarray}
Basis correlation functions for central and perimeter sites are defined respectively as follows:
\begin{eqnarray}
\nonumber
m=\left\langle\left\langle c_{0}S_{0}\right\rangle\right\rangle_{r}=x_{1}&=&(3c-6c^{2}+3c^{3})x_{4}K_{1}+(6c^{2}-6c^{3})x_{4}K_{2}+3c^{3}x_{4}K_{3}+cx_{6}K_{4},\\
\nonumber
\left\langle\left\langle c_{1}S_{1}\right\rangle\right\rangle_{r}=x_{4}&=&(c-c^{2})A_{1}+c^{2}A_{2}+cx_{1}A_{3}.
\end{eqnarray}
By expanding Eq. (\ref{eq6}) with $i=0$ and $\{f_{i}\}=c_{1}S_{1}$ we get
\begin{eqnarray}
\nonumber
\langle\langle c_{1}S_{1}c_{0}S_{0}\rangle\rangle_{r}=x_{2}=(3c^{2}-6c^{3}+3c^{4})K_{1}+(6c^{3}-6c^{4})K_{2}+3c^{4}K_{3}+c^{2}x_{5}K_{4}.
\end{eqnarray}
By the same way, putting $i=0$ and $\{f_{i}\}=c_{1}S_{1}c_{2}S_{2}$ in Eq. (\ref{eq6}) we obtain
\begin{eqnarray}
\nonumber
x_{3}=(-3c^{2}+3c^{3})x_{4}K_{1}+(6c^{2}-6c^{3})x_{4}K_{2}+3c^{3}x_{4}K_{3}+c^{3}x_{4}K_{4}.
\end{eqnarray}
Similarly, by using Eq. (\ref{eq15}) with $\delta=1$ and $\{f_{\delta}\}=c_{2}S_{2}$, and $\{f_{\delta}\}=c_{2}S_{2}c_{3}S_{3}$ we find $x_{5}$ and $x_{6}$, respectively. Hence, we get the complete set of correlation functions as follows:
\begin{eqnarray}\label{eq2a}
\nonumber
x_{1}&=&(3c-6c^{2}+3c^{3})x_{4}K_{1}+(6c^{2}-6c^{3})x_{4}K_{2}+3c^{3}x_{4}K_{3}+cx_{6}K_{4},\\
\nonumber
x_{2}&=&(3c^{2}-6c^{3}+3c^{4})K_{1}+(6c^{3}-6c^{4})K_{2}+3c^{4}K_{3}+c^{2}x_{5}K_{4},\\
\nonumber
x_{3}&=&(-3c^{2}+3c^{3})x_{4}K_{1}+(6c^{2}-6c^{3})x_{4}K_{2}+3c^{3}x_{4}K_{3}+c^{3}x_{4}K_{4},\\
\nonumber
x_{4}&=&(c-c^{2})A_{1}+c^{2}A_{2}+cx_{1}A_{3},\\
\nonumber
x_{5}&=&cx_{2}A_{3},\\
x_{6}&=&cx_{3}A_{3}.
\end{eqnarray}
where the coefficients $K_{n}, (n=1,...,4)$ and $A_{l}, (l=1,...,3)$ are given in Eqs. (\ref{eq13}) and (\ref{eq17}), respectively.

On the other hand, corresponding to Eq. (\ref{eq1a}), for a square lattice $(q=4)$ we have
\begin{eqnarray}\label{eq3a}
\nonumber
x_{1}&=&\langle\langle c_{0}S_{0}\rangle\rangle_{r},\\
\nonumber
x_{2}&=&\langle\langle c_{0}S_{0}c_{1}S_{1}\rangle\rangle_{r},\\
\nonumber
x_{3}&=&\langle\langle c_{0}S_{0}c_{1}S_{1}c_{2}S_{2}\rangle\rangle_{r},\\
\nonumber
x_{4}&=&\langle\langle c_{0}S_{0}c_{1}S_{1}c_{2}S_{2}c_{3}S_{3}\rangle\rangle_{r},\\
\nonumber
x_{5}&=&\langle\langle c_{1}S_{1}\rangle\rangle_{r},\\
\nonumber
x_{6}&=&\langle\langle c_{1}S_{1}c_{2}S_{2}\rangle\rangle_{r},\\
\nonumber
x_{7}&=&\langle\langle c_{1}S_{1}c_{2}S_{2}c_{3}S_{3}\rangle\rangle_{r},\\
x_{8}&=&\langle\langle c_{1}S_{1}c_{2}S_{2}c_{3}S_{3}c_{4}S_{4}\rangle\rangle_{r}.
\end{eqnarray}

By following the same procedure given for $q=3$ above, we get the complete set of linear equations for a square lattice $(q=4)$ as follows:
\begin{eqnarray}\label{eq4a}
\nonumber
x_{1}&=&(4c-12c^{2}+12c^{3}-4c^{4})x_{5}L_{1}+(12c^{2}-24c^{3}+12c^{4})x_{5}L_{2}+(12c^{3}-12c^{4})x_{5}L_{3}\\
\nonumber
&&+4c^{4}x_{5}L_{4}+(4c-4c^{2})x_{7}L_{5}+4c^{2}x_{7}L_{6},\\
\nonumber
x_{2}&=&(4c^{2}-12c^{3}+12c^{4}-4c^{5})L_{1}+(12c^{3}-24c^{4}+12c^{5})L_{2}+(12c^{4}-12c^{5})L_{3}+4c^{5}L_{4}\\
\nonumber
&&+(4c^{2}-4c^{3})x_{6}L_{5}+4c^{3}x_{6}L_{6},\\
\nonumber
x_{3}&=&(-8c^{2}+12c^{3}-4c^{4})x_{5}L_{1}+(12c^{2}-24c^{3}+12c^{4})x_{5}L_{2}+(12c^{3}-12c^{4})x_{5}L_{3}\\
\nonumber
&&+4c^{4}x_{5}L_{4}+(4c^{3}-4c^{4})x_{5}L_{5}+4c^{4}x_{5}L_{6},\\
\nonumber
x_{4}&=&(4c^{2}-4c^{3})x_{6}L_{1}+(-12c^{2}+12c^{3})x_{6}L_{2}+(12c^{2}-12c^{3})x_{6}L_{3}+4c^{3}x_{6}L_{4}\\
\nonumber
&&+(4c^{2}-4c^{3})x_{6}L_{5}+4c^{3}x_{6}L_{6},\\
\nonumber
x_{5}&=&(c-c^{2})B_{1}+c^{2}B_{2}+cx_{1}B_{3},\\
\nonumber
x_{6}&=&(c-c^{2})x_{5}B_{1}+c^{2}x_{5}B_{2}+cx_{2}B_{3},\\
\nonumber
x_{7}&=&(c-c^{2})x_{6}B_{1}+c^{2}x_{6}B_{2}+cx_{3}B_{3},\\
x_{8}&=&(c-c^{2})x_{7}B_{1}+c^{2}x_{7}B_{2}+cx_{4}B_{3},
\end{eqnarray}
where
\begin{eqnarray}
\nonumber
L_{1}=\sinh(J\nabla)\tanh(\beta x)|_{x=0},\ \ \ \ \ \ \ \ \ \ \ \ \ \ \ \ \ &&\\
\nonumber
L_{2}=\cosh(J\nabla)\sinh(J\nabla)\tanh(\beta x)|_{x=0},\ \ \ \ &&\quad B_{1}=\tanh(\beta(x+\gamma))|_{x=0},\\
\nonumber
L_{3}=\cosh^{2}(J\nabla)\sinh(J\nabla)\tanh(\beta x)|_{x=0},\ \ \ &&\quad B_{2}=\cosh(J\nabla)\tanh(\beta(x+\gamma))|_{x=0}, \\
\nonumber
L_{4}=\cosh^{3}(J\nabla)\sinh(J\nabla)\tanh(\beta x)|_{x=0},\ \ \ &&\quad B_{3}=\sinh(J\nabla)\tanh(\beta(x+\gamma))|_{x=0},\\
\nonumber
L_{5}=\sinh^{3}(J\nabla)\tanh(\beta x)|_{x=0}, \ \ \ \ \ \ \ \ \ \ \ \ \ \ \ \ &&\\
\nonumber
L_{6}=\cosh(J\nabla)\sinh^{3}(J\nabla)\tanh(\beta x)|_{x=0}. \ \ \ &&
\end{eqnarray}
Phase diagrams and magnetization curves can be obtained by solving Eq. (\ref{eq4a}) numerically with the condition
\begin{equation}\label{eq5a}
x_{1}=x_{5}.
\end{equation}
\section{Derivation of complete set of linear equations for spin-1 BC model}\label{appendixb}
We label the site correlations as  $x_{i}$, $i=1,2,...,21$. The complete list is as follows
\begin{eqnarray}\label{eq1b}
\nonumber
x_{1}=\langle\langle c_{0}S_{0}\rangle\rangle_{r},&& x_{12}=\langle\langle c_{0}S_{0}c_{1}S_{1}^{2}c_{2}S_{2}^{2}\rangle\rangle_{r},\\
\nonumber
x_{2}=\langle\langle c_{0}S_{0}c_{1}S_{1}\rangle\rangle_{r},&&x_{13}=\langle\langle c_{1}S_{1}c_{2}S_{2}c_{3}S_{3}^{2}\rangle\rangle_{r},\\
\nonumber
x_{3}=\langle\langle c_{0}S_{0}c_{1}S_{1}c_{2}S_{2}\rangle\rangle_{r},&&x_{13}=\langle\langle c_{1}S_{1}c_{2}S_{2}^{2}c_{3}S_{3}^{2}\rangle\rangle_{r},\\
\nonumber
x_{4}=\langle\langle c_{1}S_{1}\rangle\rangle_{r},&&x_{15}=\langle\langle c_{1}S_{1}^{2}c_{2}S_{2}^{2}c_{3}S_{3}^{2}\rangle\rangle_{r},\\
\nonumber
x_{5}=\langle\langle c_{1}S_{1}c_{2}S_{2}\rangle\rangle_{r},&&x_{16}=\langle\langle c_{0}S_{0}^{2}\rangle\rangle_{r},\\
\nonumber
x_{6}=\langle\langle c_{1}S_{1}c_{2}S_{2}c_{3}S_{3}\rangle\rangle_{r},&&x_{17}=\langle\langle c_{0}S_{0}^{2}c_{1}S_{1}\rangle\rangle_{r},\\
\nonumber
x_{7}=\langle\langle c_{1}S_{1}^{2}\rangle\rangle_{r},&&x_{18}=\langle\langle c_{0}S_{0}^{2}c_{1}S_{1}^{2}\rangle\rangle_{r},\\
\nonumber
x_{8}=\langle\langle c_{1}S_{1}c_{2}S_{2}^{2}\rangle\rangle_{r},&&x_{19}=\langle\langle c_{0}S_{0}^{2}c_{1}S_{1}c_{2}S_{2}\rangle\rangle_{r},\\
\nonumber
x_{9}=\langle\langle c_{1}S_{1}^{2}c_{2}S_{2}^{2}\rangle\rangle_{r},&&x_{20}=\langle\langle c_{0}S_{0}^{2}c_{1}S_{1}c_{2}S_{2}^{2}\rangle\rangle_{r},\\
\nonumber
x_{10}=\langle\langle c_{0}S_{0}c_{1}S_{1}^{2}\rangle\rangle_{r},&&x_{21}=\langle\langle c_{0}S_{0}^{2}c_{1}S_{1}^{2}c_{2}S_{2}^{2}\rangle\rangle_{r},\\
x_{11}=\langle\langle c_{0}S_{0}c_{1}S_{1}c_{2}S_{2}^{2}\rangle\rangle_{r}.&&
\end{eqnarray}

The correlation functions $x_{i}$, $i=1,2,3$ are obtained from Eq. (\ref{eq22}). For example,  putting $i=0$ and $\{f_{i}\}=c_{1}S_{1}$ and $\{f_{i}\}=c_{1}S_{1}c_{2}S_{2}$ in Eq. (\ref{eq22}) we obtain $x_{2}$ and $x_{3}$ correlation functions, respectively as follows
\begin{eqnarray}
\nonumber
\langle\langle c_{0}S_{0}c_{1}S_{1}\rangle\rangle_{r}=x_{2}&=&3ck_{1}x_{7}+(-6k_{1}+6k_{2})cx_{9}+ck_{3}x_{13}+(3k_{1}-6k_{2}+3k_{4})cx_{15},\\
\nonumber
\langle\langle c_{0}S_{0}c_{1}S_{1}c_{2}S_{2}\rangle\rangle_{r}=x_{3}&=&(-3k_{1}+6k_{2})cx_{8}+(3k_{1}-6k_{2}+k_{3}+3k_{4})cx_{14},
\end{eqnarray}
The equations labeled $x_{j}$ with $j=4,5,6$ are derived from Eq. (\ref{eq29}). In a similar way, the correlation functions $x_{k}$ with $k=7,8,...,15$ and $x_{l}$ with $l=16,17,...,21$ can be easily obtained by using Eqs. (\ref{eq23}) and (\ref{eq30}), respectively. By following the above procedure, we can get the complete set of linear equations as follows:

\begin{eqnarray}\label{eq2b}
\nonumber
x_{1}&=&3cx_{4}k_{1}+cx_{6}k_{3}+(-6k_{1}+6k_{2})cx_{8}+(3k_{1}-6k_{2}+3k_{4})cx_{14},\\
\nonumber
x_{2}&=&3ck_{1}x_{7}+(-6k_{1}+6k_{2})cx_{9}+ck_{3}x_{13}+(3k_{1}-6k_{2}+3k_{4})cx_{15},\\
\nonumber
x_{3}&=&(-3k_{1}+6k_{2})cx_{8}+(3k_{1}-6k_{2}+k_{3}+3k_{4})cx_{14},\\
\nonumber
x_{4}&=&a_{1}c+a_{2}cx_{1}+(a_{3}-a_{1})cx_{16},\\
\nonumber
x_{5}&=&a_{1}cx_{4}+a_{2}cx_{2}+(a_{3}-a_{1})cx_{17},\\
\nonumber
x_{6}&=&a_{1}cx_{5}+a_{2}cx_{3}+(a_{3}-a_{1})cx_{19},\\
\nonumber
x_{7}&=&b_{1}c+b_{2}cx_{1}+(b_{3}-b_{1})cx_{16},\\
\nonumber
x_{8}&=&b_{1}cx_{4}+b_{2}cx_{2}+(b_{3}-b_{1})cx_{17},\\
\nonumber
x_{9}&=&b_{1}cx_{7}+b_{2}cx_{10}+(b_{3}-b_{1})cx_{18},\\
\nonumber
x_{10}&=&b_{2}cx_{16}+b_{3}cx_{1},\\
\nonumber
x_{11}&=&b_{2}cx_{17}+b_{3}cx_{2},\\
\nonumber
x_{12}&=&b_{2}cx_{18}+b_{3}cx_{10},\\
\nonumber
x_{13}&=&b_{1}cx_{5}+b_{2}cx_{3}+(b_{3}-b_{1})cx_{19},\\
\nonumber
x_{14}&=&b_{1}cx_{8}+b_{2}cx_{11}+(b_{3}-b_{1})cx_{20},\\
\nonumber
x_{15}&=&b_{1}cx_{9}+b_{2}cx_{12}+(b_{3}-b_{1})cx_{21},\\
\nonumber
x_{16}&=&cr_{0}+3cr_{2}x_{5}+(-3r_{0}+3r_{1})cx_{7}+(3r_{0}-6r_{1}+3r_{3})x_{9}\\
\nonumber
&&+(-3r_{2}+3r_{4})cx_{13}+(-r_{0}+3r_{1}-3r_{3}+r_{5})cx_{15},\\
\nonumber
x_{17}&=&(-2r_{0}+3r_{1})cx_{4}+(3r_{0}+3r_{2}+3r_{3}-6r_{1})cx_{8}\\
\nonumber
&&+(-r_{0}+3r_{1}-3r_{2}-3r_{3}+3r_{4}+r_{5})cx_{14},\\
\nonumber
x_{18}&=&(-2r_{0}+3r_{1})cx_{7}+(3r_{0}-6r_{1}+3r_{2}+3r_{3})cx_{9}\\
\nonumber
&&+(-r_{0}+3r_{1}-3r_{2}-3r_{3}+3r_{4}+r_{5})x_{15},\\
\nonumber
x_{19}&=&(r_{0}-3r_{1}+3r_{2}+3r_{3})cx_{5}\\
\nonumber
&&+(-r_{0}+3r_{1}-3r_{2}-3r_{3}+3r_{4}+r_{5})cx_{13},\\
\nonumber
x_{20}&=&(r_{0}-3r_{1}+3r_{2}+3r_{3})cx_{8}\\
\nonumber
&&+(-r_{0}+3r_{1}-3r_{2}-3r_{3}+3r_{4}+r_{5})cx_{14},\\
\nonumber
x_{21}&=&(r_{0}-3r_{1}+3r_{2}+3r_{3})cx_{9}\\
&&+(-r_{0}+3r_{1}-3r_{2}-3r_{3}+3r_{4}+r_{5})cx_{15}.
\end{eqnarray}

\newpage

\end{document}